\pdfoutput=1
\documentclass[%
 reprint, amsmath, amssymb, aps,]{revtex4-1}

\usepackage{graphicx}% Include figure files
\usepackage{dcolumn}% Align table columns on decimal point
\usepackage{bm}% bold math
\usepackage{color}

\newcommand{\RN}[1]{%
  \textup{\uppercase\expandafter{\romannumeral#1}}%
}

\begin{document}

\preprint{APS/123-QED}

\title{Ultrahigh-energy cosmic-ray nuclei and neutrinos from engine-driven supernovae}

\author{B. Theodore Zhang$^{1,2}$}
\author{Kohta Murase$^{3,4,5,6}$}

\affiliation{$^1$Department of Astronomy, School of Physics, Peking University, Beijing 100871, China }
\affiliation{$^2$Kavli Institute for Astronomy and Astrophysics, Peking University, Beijing 100871, China}
\affiliation{$^3$Department of Physics, The Pennsylvania State University, University Park, Pennsylvania 16802, USA}
\affiliation{$^4$Department of Astronomy \& Astrophysics, The Pennsylvania State University, University Park, Pennsylvania 16802, USA}
\affiliation{$^5$Center for Particle and Gravitational Astrophysics, The Pennsylvania State University, University Park, Pennsylvania 16802, USA}
\affiliation{$^6$Yukawa Institute for Theoretical Physics, Kyoto, Kyoto 606-8502 Japan}

\date{\today}% It is always \today, today,
             
\begin{abstract}
Transrelativistic supernovae (SNe), which are likely driven by central engines via jets or winds, have been among candidate sources of ultrahigh-energy cosmic rays (UHECRs).  We investigate acceleration and survival of UHECR nuclei in the external reverse shock scenario. With composition models used in Zhang et al. (2018), we calculate spectra of escaping cosmic rays and secondary neutrinos. If their local rate is $\sim1$\% of the core-collapse supernova rate, the observed UHECR spectrum and composition can be explained with the total cosmic-ray energy ${\mathcal E}_{\rm cr}\sim10^{51}$~erg. The maximum energy of UHECR nuclei can reach $\sim 10^{20}-{10}^{21}\rm~eV$. 
The diffuse flux of source neutrinos is predicted to be $\sim 10^{-11} - 10^{-10}~{\rm GeV}~{\rm cm}^{-2}~{\rm s}^{-1}~{\rm sr}^{-1}$ in the 0.1-1~EeV range, satisfying nucleus-survival bounds. The associated cosmogenic neutrino flux is calculated, and shown to be comparable or even higher than the source neutrino flux.  These ultrahigh-energy neutrinos can be detected by ultimate detectors such as the Giant Radio Askaryan Neutrino Detector and Probe Of Extreme Multi-Messenger Astrophysics.
\end{abstract}

\pacs{Valid PACS appear here}% PACS, the Physics and Astronomy
\maketitle

\section{\label{sec:1}Introduction}
The deaths of massive stars lead to the production of cosmic rays, which have been established by cumulative evidences from studies on supernova (SN) remnants. Ordinary SNe are accompanied by a nonrelativistic shock, at which the diffusive shock acceleration occurs. They could be responsible for observed cosmic rays up to the knee at $\sim3\times{10}^{15}$~eV or iron knee around $10^{17}$~eV, but the maximum energy cannot reach ultrahigh energies. Among various candidate sources of ultrahigh-energy cosmic rays (UHECRs)~\cite{Hillas:1985is,Kotera:2011cp, Anchordoqui:2018qom}, rarer but faster SNe have been suggested by various authors~\cite{Murase:2006mm, Wang:2007xj, Budnik:2007yh, Wang:2007ya, Murase:2008mr,  2009PhRvD..79j3001M,Chakraborti:2010ha, Liu:2011cua, Liu:2011tv, Zhang:2017moz} (see Ref.~\cite{Murase:2008sa} for a list of transient sources for UHECRs).
Indeed, observations have revealed that some SNe have a transrelativistic component with shock velocity $\beta_s\sim0.1-1$. They include engine-driven SNe associated with low-luminosity (LL) gamma-ray bursts (GRBs), such as GRB 980425 with SN 1998bw~\cite{Galama:1998ea, Kulkarni:1998qk}, GRB 060218 with SN 2006aj~\cite{Soderberg:2006vh, Campana:2006qe}, GRB 100316D with SN 2010bh~\cite{Margutti:2013pra}, GRB 161228B with iPTF17cw~\cite{Corsi:2017hdx}, and GRB 171205A with SN 2017iuk~\cite{DElia:2018xrz, Wang:2018wqx, Izzo:2019akc}. 
Such engine-driven SNe with transrelativistic ejecta have been found through radio observations; examples without GRB counterparts include SN 2009bb~\cite{Soderberg:2009ps} (but see also Ref.~\cite{2015ApJ...805..164N}) and SN 2012ap~\cite{Chakraborti:2014dha, Margutti:2014gha}. 
Another possible example with low-mass subrelativistic ejecta is AT2018cow, although its origin is under debate~\cite{Perley:2018oky,Margutti:2018rri}.  
It has been shown that the event rate of SN 2009bb-like SNe is about 1\% of the core-collapse SN rate, which is not far from the true rate of LL GRBs, $\sim100 - 1000 \rm~Gpc^{-3}~yr^{-1}$~\cite{Chakraborti:2010ha, Sun:2015bda}. It is also comparable to the hypernova rate~\cite{2004ApJ...607L..17P,2007ApJ...657L..73G}, although all hypernovae are not necessarily accompanied by transrelativistic ejecta. 
The diversity of these explosive phenomena may originate from characteristics of central engines and/or progenitors~\cite{2004ApJ...611..380T}, and the relativistic velocity component can be caused by energy injections via outflows from the engines~\cite{Lazzati:2011ay,2015ApJ...807..172N,2017ApJ...834...32S}. 

The diffusive shock acceleration theory predicts that the acceleration time scale is $\sim E/(ZeBc\beta_s^2)$, where $E$ is particle energy, $Z$ is the charge, and $B$ is magnetic field strength. We can see the acceleration is easier for faster shocks ($\Gamma \gg 1$), but it is known that ultrarelativistic, superluminal shocks are unlikely to be efficient cosmic-ray accelerators~\cite{2001MNRAS.328..393A}. 
In this sense, engine-driven SNe, whose shocks are subrelativistic or mildly relativistic, $\Gamma \lesssim 1-10$, would be more favorable for the acceleration of UHECR nuclei. 
The magnetic field strength inferred from radio observations~\cite{Chakraborti:2010ha} also supports that the acceleration of UHECR nuclei is allowed by the Hillas criterion~\cite{Hillas:1985is}.
It has been shown that the composition of UHECR nuclei becomes heavier beyond ``ankle'' ($\sim 4 \times 10^{18}\rm~eV$)~\cite{Aab:2017njo} which requires their sources to have super-solar abundance~\cite{Aab:2016zth}. 
Contrary to the forward shock (FS) scenario, in the internal shock and reverse shock (RS) scenarios, an outflow or ejecta composition with intermediate mass nuclei enriched is naturally achieved in terms of progenitor models~\cite{Zhang:2017moz}. 

If UHECRs are dominated by intermediate or heavy nuclei, source identification with UHECR observations is more challenging because of larger magnetic deflections. The requirement of nucleus survival also restricts the production of high-energy neutrinos both inside and outside the sources~\cite{Murase:2010gj}. Detections of neutrinos or electromagnetic counterparts from the UHECR accelerators will provide us with a unique opportunity to reveal the sources, as studied for LL GRBs and transrelativistic SNe~\cite{Murase:2006mm,Gupta:2006jm,Murase:2008mr,Murase:2009ah,Murase:2010va,Liu:2011cua,2012ApJ...745L..16M,2013ApJ...769L...6K,Murase:2013ffa,Senno:2015tsn,Boncioli:2018lrv,Denton:2018tdj}.
Other source classes including galaxy clusters~\cite{Murase:2008yt, Fang:2017zjf}, activate galactic nuclei~\cite{Murase:2015ndr}, classical GRBs~\cite{Waxman:1995vg, Murase:2008mr, Globus:2014fka, Biehl:2017zlw}, tidal disruption events (TDEs)~\cite{Zhang:2017hom, Biehl:2017hnb, Guepin:2017abw}, and new born pulsars~\cite{Fang:2013vla}.
Neutrinos that directly originate from UHECR nuclei have energy with $E_\nu\approx0.05(E/A)\simeq 0.5~{\rm EeV}~{(E/{10}^{20}~\rm eV)}(10/A)$. This energy range is higher than that of IceCube neutrinos, 0.1-1~PeV. 
Such extremely high-energy neutrinos are the main targets for planned neutrino detectors in the near future, Askaryan Radio Array (ARA)~\cite{Allison:2015eky}, Antarctic Ross Ice Shelf Antenna Neutrino Array (ARIANNA)~\cite{Barwick:2014pca}, Giant Radio Array for Neutrino Detection (GRAND)~\cite{Alvarez-Muniz:2018bhp}, Probe Of Extreme Multi-Messenger Astrophysics (POEMMA)~\cite{Olinto:2017xbi}, and Trinity~\cite{Otte:2018uxj}. 
The discovery of IceCube-170922A, coinciding with a flaring blazar, TXS 0506+056, demonstrated the feasibility of neutrino-triggered follow-up observations at different wavelengths~\cite{IceCube:2018dnn,Keivani:2018rnh,Ahnen:2018mvi}, although there is no simple picture for the 2017 and 2014-2015 flares~\cite{Murase:2018iyl}. LL GRBs and engine-driven SNe will be interesting targets for such follow-up observations, as proposed by Ref.~\cite{Murase:2006mm}. 

In addition to the source neutrinos, cosmogenic neutrinos which are produced during the propagation of UHECR nuclei in the intergalactic space are believed to be guaranteed possibilities, even though the expected neutrino flux are subject to $\sim 1 - 2$ orders of magnitude uncertainty depending on the composition, maximum acceleration energy and source redshift evolution~\cite{Beresinsky:1969qj, Stecker:1978ah, Engel:2001hd, Takami:2007pp,Kotera:2010yn, AlvesBatista:2018zui}. 

In this work, we provide a comprehensive study of the generation and survival of UHECR nuclei and neutrinos from engine-driven SNe. 
For general consideration, we discuss two kinds of outflows; one is mildly relativistic jets with $\Gamma\sim2-10$, and the other is the slower transrelativistic ejecta with $\Gamma \beta \sim 1$. 
We revisit the ``nucleus-survival problem'' that was earlier studied by Ref.~\cite{Murase:2008mr,Wang:2007xj,Murase:2010gj}, and then calculate spectra of escaping cosmic rays and secondary neutrinos both numerically and analytically. Our results demonstrate that engine-driven SNe accompanied by LL GRBs jets~\cite{Murase:2008mr} or transrelativistic ejecta~\cite{Liu:2011tv} give a viable explanation for the UHECR data even in the RS scenario. 

This paper is organized as follows. 
We first study the physics of RS formed by both jets and transrelativistic ejecta in Sec.~\ref{sec:two}.
In Sec.~\ref{sec:three}, we discuss the acceleration of UHECR nuclei in the RS scenario and compared to the observation results of UHECR nuclei.
In Sec.~\ref{sec:four}, we estimate neutrinos that are coproduced with UHECR nuclei from jets or winds considering the photomeson production process between UHECR nuclei and ambient radiation fields. We concentrate on neutrino fluences from single source and diffuse neutrinos which takes into account the contribution from all of the events in the universe and cosmogenic neutrinos.
We discuss implications of our results in Sec.~\ref{sec:five} and give a summary in Sec.~\ref{sec:six}.

Throughout of the paper, we adopt the cgs unit and have notations $Q_x \equiv Q /10^x$. The cosmological parameters we use are $H_0 = 67.3 \ \rm km \ s^{-1} \ Mpc^{-1}$, $\Omega_m = 0.315$, $\Omega_\Lambda = 0.685$ \cite{Agashe:2014kda}. 

\section{\label{sec:two}The physics of reverse shock}
\begin{figure}
\includegraphics[width=\linewidth]{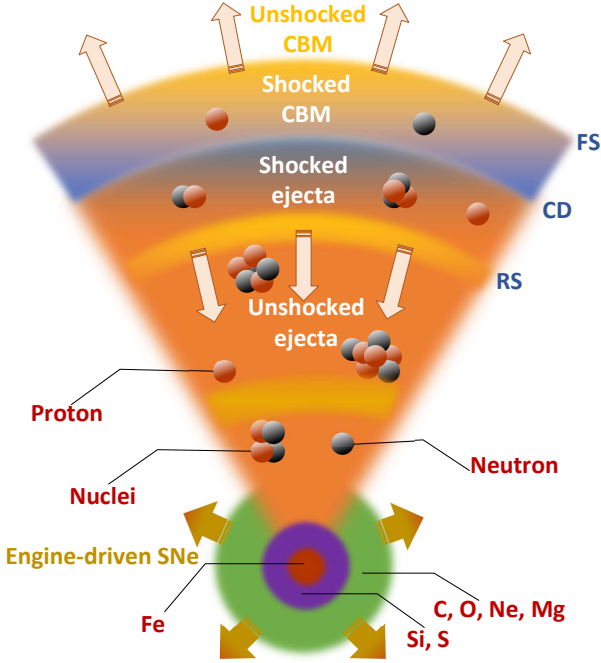}
\caption{A schematic diagram about the origin of UHECR nuclei from GRBs. Nuclei in the stellar core can be extracted by the relativistic outflow and accelerated to ultrahigh energies in the energy dissipation region via internal shocks or external reverse-forward shocks. The progenitor massive star is assumed to have an ``onion-skin'' structure at the onset of core collapse with an iron core in the center (red circle) surrounded by Silicon/Sulfur shell (purple circle) and Oxygen/Carbon shell (green circle). \label{fig:GRB}}
\end{figure}

After the core collapse of a progenitor star, outflows from the engine, either in the form of jets or winds, will push the ejecta. We expect that such engine-driven ejecta posses an enhanced fast-velocity component~\cite{Izzo:2019akc}, becoming transrelativistic SNe. The material collides into the surrounding circumburst medium (CBM), leading to the formation of external reverse-forward shocks~\cite{2006RPPh...69.2259M}. 
%At the deceleration radius, the total energy of the swept-up CBM is equal to the kinetic energy of the ejecta. 
While the FS plows the CBM directly, the RS propagates back and decelerate the outflow. Most of the RS emission occurs around the shell crossing time $t_{\times}$. If the engine duration $T_{\rm eng}$ is longer than the deceleration time $t_{\rm dec}$, corresponding to the thick shell case, the shell crossing time is approximately given by $\sim T_{\rm eng}$. In more general, it is given by $t_{\times}\sim{\rm max}[T_{\rm eng},t_{\rm dec}]$. 
The shocked ejecta and shocked CBM are separated by the contact discontinuity (CD) where the pressure equilibrium has been established across the interacting surface, see Fig.~\ref{fig:GRB}. 

One of the particular features of models for acceleration by engine-driven outflows is that they may be composed by a large fraction of heavy nuclei as we have illustrated in Fig.~\ref{fig:GRB}.  Heavy nuclei can be extracted from the inner stellar core~\cite{Horiuchi:2012by, Zhang:2017moz} or synthesized during the expansion of the outflow~\cite{Nakamura:2000ms, Liu:2011tv,2011MNRAS.415.2495M}. 
In this work, we adopt the nuclear composition model, Si-R \RN{1}, proposed in Ref.~\cite{Zhang:2017moz} as a fiducial model of the jet composition, while the transrelativistic ejecta has a composition similar to the hypernova model in Ref.~\cite{Zhang:2017moz}. 

\subsection{\label{sec:two_A}Reverse shock by jets}
An observed signature of RS emission is the early optical or radio afterglow at the end of the main burst of GRBs~\cite{Akerlof:1999aa, Laskar:2013uza, Laskar:2018tnq}. The RS emission is important to constrain the initial Lorentz factor as well as the baryonic component in the jets~\cite{Zhang:2003wj, Nakar:2004wf, McMahon:2005je}. The RS emission can be suppressed or even missed if the ejecta is dominated by the magnetic energy, but we can expect strong RS emission for baryon dominated jets~\cite{Zhang:2004ie, Kumar:2014upa}, which are promising for engine-driven SNe.

The dynamical properties of RS can be more complicated than the self-similar evolution of FS, especially in the presence of long-lasting energy injection~\cite{2007ApJ...665L..93U,2007MNRAS.381..732G}. 
For our purpose, we only consider the shock physics at radius $R_\times$, where RS finishes crossing the GRB ejecta shell~\cite{Panaitescu:2004sn, Murase:2007yt}. After the shock crossing, the RS light curve is expected to decline rapidly with time $\propto t^{-2}$ as a result of the rapid cooling of electrons~\cite{Zhang:2003wj, Kobayashi:2003zk}. 
In the following, our calculations are based on the work in Ref.~\cite{Murase:2007yt}.  

We assume that the GRB ejecta have an initial Lorentz factor $\Gamma_0$ and isotropic-equivalent ejecta energy $\mathcal{E}_{\rm k}$. The CBM density is given by $\varrho_{\rm cbm}$, and $\varrho_{\rm ej}$ is the number density of the GRB ejecta, which can be estimated by $\varrho_{\rm ej} = \mathcal{E}_{\rm k} / 4 \pi m_p c^2 \Gamma_0 (\Gamma_0 \Delta) R^2$ with $\Delta$ as the ejecta shell thickness in the engine frame. 
For a homogeneous CBM with $\varrho_{\rm cbm} = \rm const$, the RS crossing radius $R_\times$ is estimated to be $R_\times \simeq 5.6 \times 10^{16} \mathcal{E}_{{\rm k}, 51.5}^{1/4} \varrho_{\rm cbm, 1}^{-1/4} T_{4}^{1/4} {\rm~cm}$, where we adopt the ``thick ejecta shell'' case considering $\Delta = c T > R_\times / 2 \Gamma_0^2$, and $T = 10^4 \rm~s$ is the engine frame duration of the GRB ejecta~\cite{Panaitescu:2004sn}. 
This is justified {\it when the central engine is active for a sufficiently long time.}
Note that if $R_\times / 2 \Gamma_0^2 > cT$, we should consider the ``thin ejecta shell'' $\Delta = R_\times / 2 \Gamma_0^2$, where the thickness of the ejecta shell are dominated by the velocity spreading. 

The Lorentz factor of the shocked ejecta in the engine frame is $\Gamma_\times \simeq 6.3~\mathcal{E}_{{\rm k}, 51.5}^{1/8} \varrho_{\rm cbm, 1}^{-1/8} T_{4}^{-3/8}$, where we adopt the condition $\varrho_{\rm ej} / \varrho_{\rm cbm} \ll 4 \Gamma_0^2$ for more tenuous ejecta. The Lorentz factor of the shocked ejecta viewed from the frame of the unshocked ejecta can be calculated from the addition of velocities in special relativity,  $\Gamma_\times^\prime \approx (1/2) (\Gamma_\times / \Gamma_0 + \Gamma_0 / \Gamma_\times) \simeq 1.1$. The magnetic field strength of the shocked GRB ejecta can be estimated assuming a fraction $\epsilon_B$ of the post-shock energy density is converted into the magnetic energy, $B_\times \simeq 1.6 \epsilon_{B, -1.3}^{1/2} \mathcal{E}_{{\rm k}, 51.5}^{1/8} \varrho_{\rm cbm, 1}^{3/8} T_{4}^{-3/8} {\rm~G}$.

Once we know the Lorentz factor and magnetic field strength of the shocked ejecta, we can constrain the RS emission spectra. The typical break frequencies measured in the engine frame can be calculated using the formula $\nu_i = 3 e \gamma_i^2 B_\times \Gamma_\times / 4 \pi m_e c $
with some characteristic Lorentz factor of electrons, $\gamma_i$.  Here $\nu_i$ represents $\nu_m$ (injection frequency), $\nu_a$ (self-absorption frequency), and $\nu_c$ (cooling frequency), respectively. The injection synchrotron frequency in the engine frame is
\begin{eqnarray}
\nu_m &\simeq& 1.4 \times 10^{13} [(\Gamma_\times^\prime - 1) / 0.1]^2  \nonumber \\ &\times& \epsilon_{e, -1}^2 f_{e, -2}^{-2} \epsilon_{B,-1.3}^{1/2} \mathcal{E}_{{\rm k}, 51.5}^{1/4} \varrho_{\rm cbm, 1}^{1/4} T_{4}^{-3/4}\rm~Hz,
\end{eqnarray}
with $\epsilon_e$ is the equipartition value of the thermal energy convert to electrons, $f_e$ is the number fraction of electrons that are accelerated. We adopt $s = 2.4$ as the default electron spectral index as in Ref.~\cite{Murase:2007yt}, and the chosen value $s = 2.4$ is already used in previous works in order to reproduce the external reverse-forward shock emission~\cite{Meszaros:1999kv, Panaitescu:2004sn}.
The electron cooling Lorentz factor depends on the ratio between electron radiation time scale and dynamical time scale $\gamma_c = 6\pi m_e c^2 \Gamma_\times / \sigma_T (Y+1)R_\times B_\times^2$, where $Y$ is the Compton Y parameter. 
The typical cooling frequency in the slow cooling regime is 
\begin{equation}
\nu_c \simeq 4.1 \times 10^{13}  \epsilon_{B,-1.3}^{-3/2} \mathcal{E}_{{\rm k}, 51.5}^{-1/2} \varrho_{\rm cbm, 1}^{-1} T_{4}^{-1/2}\rm~Hz,
\end{equation} 
and the self-absorption frequency is
\begin{eqnarray}
\nu_a &\simeq& 3.8 \times 10^{9} \epsilon_{B,-1.3}^{1/5} \epsilon_{e, -1}^{-1} f_{e, -2}^{8/5} \mathcal{E}_{{\rm k}, 51.5}^{19/40} \nonumber \\ &\times&  \varrho_{\rm cbm, 1}^{13/40} T_{4}^{-33/40} [(\Gamma_\times^\prime - 1) / 0.1]^{-1}\rm~Hz.
\end{eqnarray}
The latter is estimated by setting the self-absorption optical depth $\tau (\nu_a)$ to unity~\cite{Panaitescu:2004sn, Murase:2007yt}. 

\begin{figure}
\includegraphics[width=\linewidth]{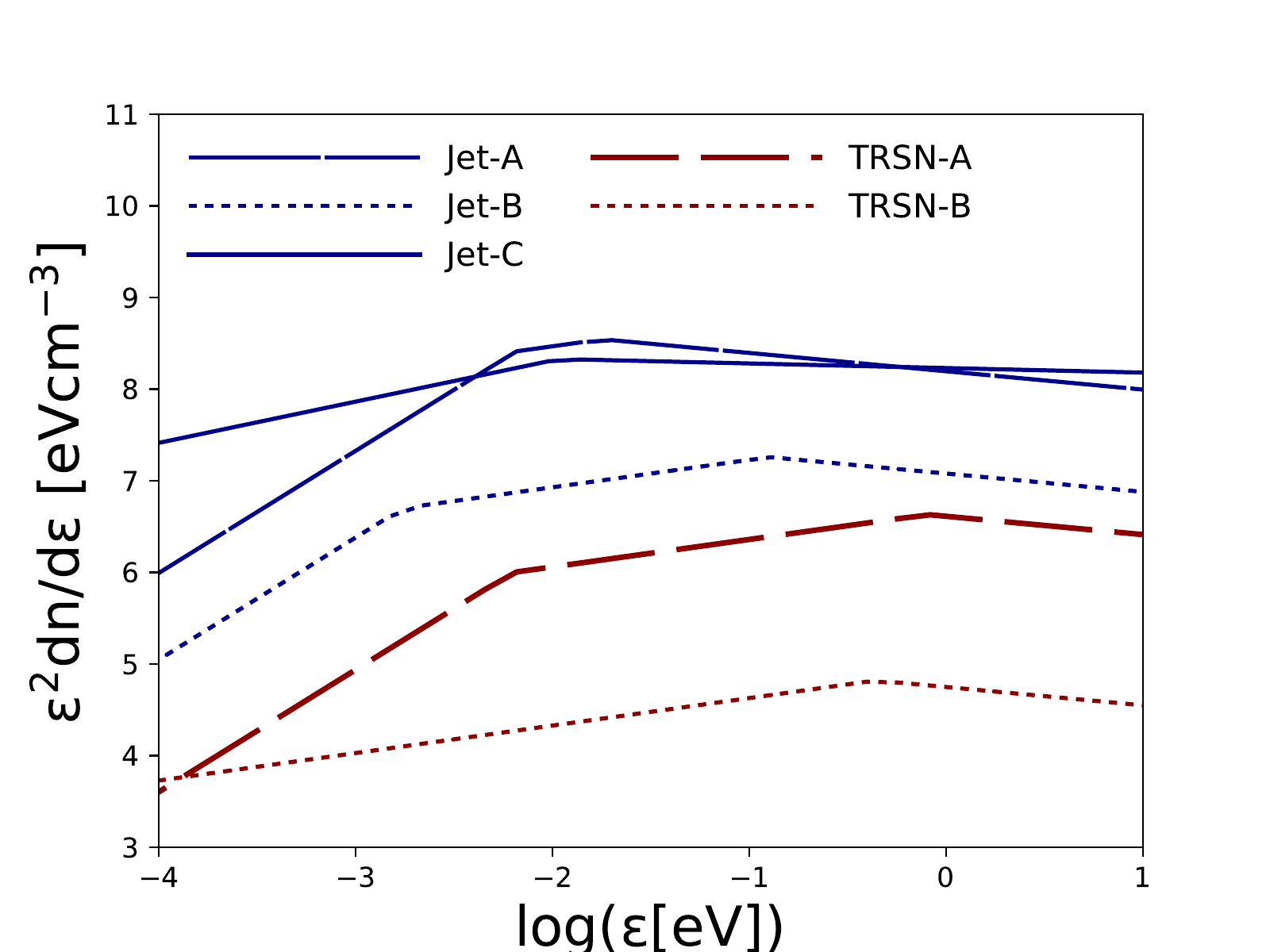}
\caption{The comoving frame differential photon number density for RS in different models: Jet-A (blue thick line), Jet-B (blue thin line), TRSN-A (red thick line), TRSN-B (red thin line).  \label{fig:RS_photon_density}}
\end{figure}

The synchrotron emission from RS can be described as broken power law~\cite{Murase:2007yt} ($\nu_a < \nu_m < \nu_c$)
\begin{eqnarray}
\frac{dn}{d\varepsilon} &=& n_{\varepsilon, \rm~max} \left\{ \begin{array}{ll} (\varepsilon_a / \varepsilon_m)^{-2/3} (\varepsilon / \varepsilon_a) & \varepsilon_{\rm min} < \varepsilon \leq \varepsilon_a \\ (\varepsilon / \varepsilon_m)^{-2/3} & \varepsilon_a < \varepsilon \leq \varepsilon_m \\ (\varepsilon / \varepsilon_m)^{-(s+1)/2} & \varepsilon_m < \varepsilon \leq \varepsilon_c \\ (\varepsilon_c / \varepsilon_m)^{-(s+1)/2} (\varepsilon / \varepsilon_c)^{-(s+2)/2} & \varepsilon_c < \varepsilon \leq \varepsilon_{\rm max} \end{array} \right.
\end{eqnarray}
where $n_{\varepsilon, \rm~max} = L_{\varepsilon, \rm~max}/4\pi R_\times^2 c \varepsilon_m$
is the normalization of the differential photon number density.
The comoving frame luminosity per unit energy is
\begin{eqnarray}
L_{\varepsilon, \rm~max} &=& \frac{1}{2\pi\hbar} \frac{f_e N_e \sqrt{3} e^3 B_\times}{m_e c^2} \nonumber \\ &=& 6.9 \times 10^{55} f_{e, -2} \epsilon_{B,-1.3}^{1/2} \mathcal{E}_{{\rm k}, 51.5}^{9/8} \varrho_{\rm cbm, 1}^{3/8} T_{4}^{-3/8} {\rm~s^{-1}},
\end{eqnarray}
where $N_e = \mathcal{E}_{\rm k} / \Gamma_0 m_p c^2$. 
We show the comoving frame differential photon number density (blue lines) in Fig.~\ref{fig:RS_photon_density}, which are calculated from following different parameter sets:
\begin{itemize}
\item Jet-A: $\mathcal{E}_{\rm k} = 3 \times 10^{51} \rm~erg$, $T = 10^4 \rm~s$, $\Gamma_0 = 10$, $\varrho_{\rm cbm} = 10 \rm~cm^{-3}$, $\epsilon_e = 0.1$, $f_e = 0.01$, $\epsilon_B = 0.01$, and $s = 2.4$.
\item Jet-B: $\mathcal{E}_{\rm k} = 3 \times 10^{51} \rm~erg$, $T = 10^4 \rm~s$, $\Gamma_0 = 10$, $\varrho_{\rm cbm} = 1 \rm~cm^{-3}$, $\epsilon_e = 0.1$, $f_e = 0.01$, $\epsilon_B = 0.01$, and $s = 2.4$.
\item Jet-C: $\mathcal{E}_{\rm k} = 4 \times 10^{51} \rm~erg$, $T = 10^4 \rm~s$, $\Gamma_0 = 5$, $\varrho_{\rm cbm} = 10000 \rm~cm^{-3}$, $\epsilon_e = 0.001$, $f_e = 0.01$, $\epsilon_B = 0.001$, and $s = 2.1$.
\end{itemize}
%For all three models, we fixed the isotropic-equivalent ejecta kinetic energy to $\mathcal{E}_{\rm k} = \btz{3} \times 10^{51}\rm~erg$ and the duration of the GRB ejecta $T = 10^4\rm~s$.
%We assume the initial Lorentz factor of the mildly-relativistic jets are $\Gamma = 10$. %(Jet-A and Jet-B) and $\Gamma = 5$ (Jet-C). 
The detailed value of $\epsilon_e$, $f_e$, and $\epsilon_B$ depends on the microphysics of the collisionless shocks, and which are still unclear from first principles~\cite{Eichler:2005ug}. The main difference of Jet-A and Jet-B are the density of external medium. We also consider model Jet-C, which can give a well-fit to the radio afterglow of GRB 060218~\cite{Toma:2006iu}, as shown in Fig.~\ref{fig:Jet-C}. We also note that our model does not overshoot either the optical or x-ray data observed for GRB 060218.

%\btz{In Fig.~\ref{fig:RS_photon_flux}, we show the observed flux of RS emission assuming the engine-driven SNe are located at the same redshift as low-luminosity GRB 060218~\cite{Soderberg:2006vh, Campana:2006qe}. } 
%We can see that our models are consistent with the observed optical flux, especially for model Jet-C, which predict much lower observation flux.
%We also show the expected flux of FS emission at the distance $R_\times$ where RS finishes crossing the ejecta. We can see both the flux from RS and FS emission is comparable to the observational optical-UV data of GRB 060218.}

%We can see the RS emission is compatible with the observed flux at optical-UV band~\cite{Ghisellini:2006ng}, especially for model Jet-B.}

\begin{figure}
\includegraphics[width=\linewidth]{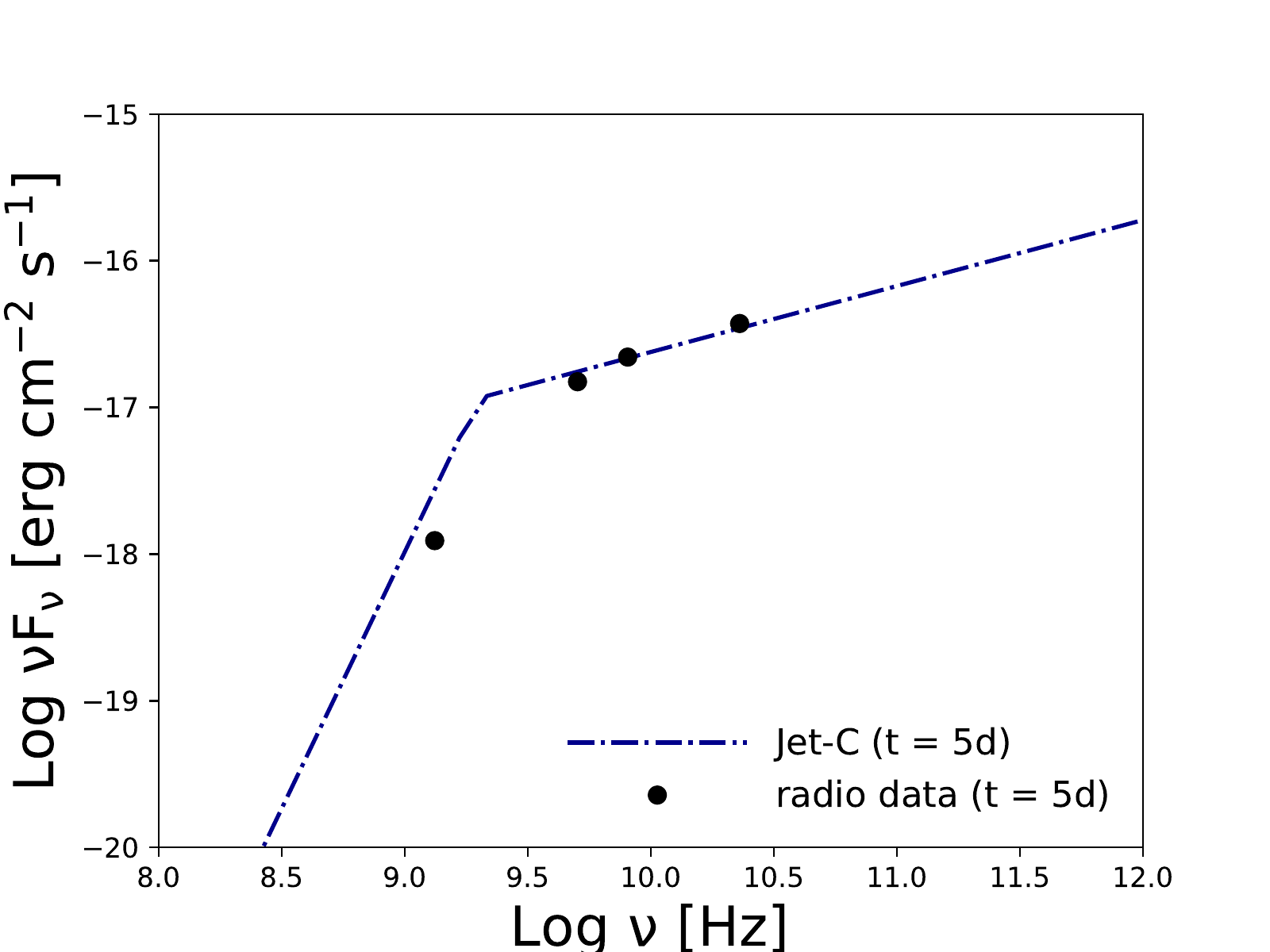}
\caption{The fitting to the radio afterglow of GRB 060218 measured at $t \sim 5\rm~d$ for model Jet-C. The radio data are taken from Ref.~\cite{Soderberg:2006vh}.
\label{fig:synchrotron_flux_update}}
\end{figure}

\begin{figure}
\includegraphics[width=\linewidth]{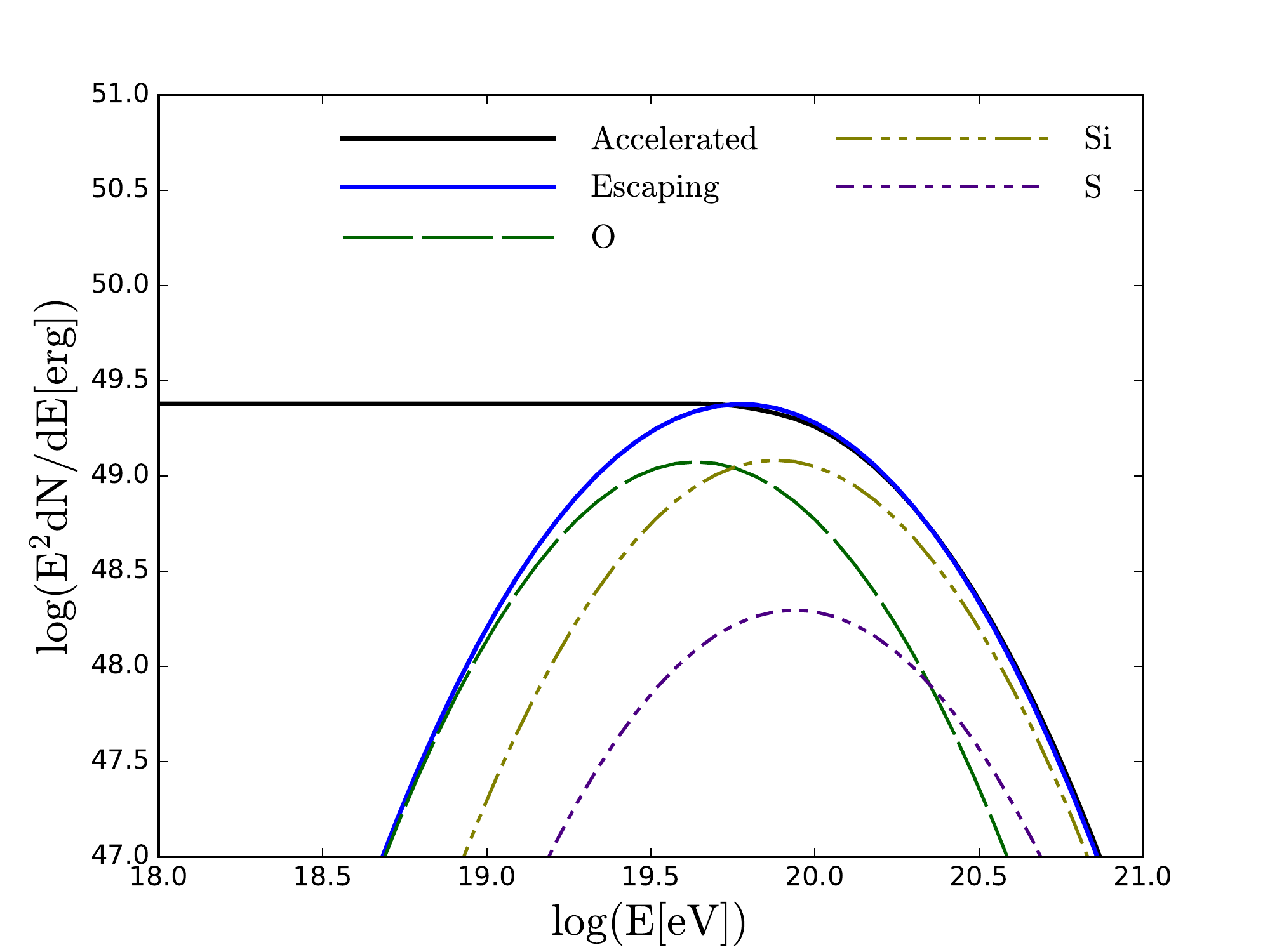}
\caption{The energy spectrum of accelerated (solid black line) and escaping (solid blue line) UHECR nuclei from engine-driven SNe for the composition model Si-R \RN{1}. The spectral index of accelerated UHECR nuclei is $s_{\rm acc} = 2$. The escaping UHECR nuclei calculated in the escape-limited scenario show a shallower cutoff than the conventional exponential cutoff. The maximum escape energy is $E_{p, \rm max}^{\rm esc} = 10^{18.3} \rm~eV$.
\label{fig:source_spectrum_cr}}
\end{figure}

\subsection{\label{sec:two_B}Reverse shock by transrelativistic ejecta}
The origin of transrelativistic SNe has been under debate but a natural possibility is that they originate from outflows or jets. 
The distribution of kinetic energy has a velocity dependence, $\mathcal{E}_{\rm ej}(>\Gamma_{\rm ej}\beta_{\rm ej}) \propto (\Gamma_{\rm ej} \beta_{\rm ej})^{-\alpha}$, where $\beta_{\rm ej} = v_{\rm ej}/c$ is the normalized velocity and $\Gamma_{\rm ej} = 1/ (1 - \beta_{\rm ej}^2)^{1/2}$ is the Lorentz factor. 
The profile is usually steep $\alpha \sim 5$ for ordinary SNe~\cite{Tan:2000nz}, whereas GRBs usually have a much flatter profile with $\alpha \simeq 0.4$~\cite{Lazzati:2011ay}. 
SNe associated with GRBs, including LL GRBs, have intermediate properties with $\alpha \sim 2-3$. This implies that most of the kinetic energy is taken by low-velocity ejecta of these SNe~\cite{Margutti:2013pra}.
For example, let us consider hypernovae with a total kinetic energy of ${\mathcal E}_{\rm ej}\sim3\times10^{52}$~erg and a typical velocity of $\beta_{\rm ej}\sim0.1$ (corresponding to an ejecta mass of $M_{\rm ej}\sim3~M_{\rm sun}$). Taking into account weights of the ejecta velocity distribution, the fast ejecta component has a typical velocity of $\beta_{\rm fej}\sim0.3$ with its characteristic energy of ${\mathcal E}_{\rm fej}\sim3\times{10}^{51}$~erg. 
In reality, the faster ejecta are decelerated earlier by the CBM. However, for simplicity, we treat the fast velocity ejecta as a single velocity component, which is sufficient for the demonstrative purpose of this work. 

The CBM is assumed to have a wind-like density profile, $\varrho_{\rm cbm} = A r^{-2}$, where $r$ is the radial distance, $A \equiv 3 \times 10^{35} A_\star {\rm~cm}^{-1}$, and $A_\star$ is the mass-loss rate to wind speed ratio normalized to $10^{-5} \rm~M_\odot~yr^{-1} / 10^3 \rm~km~s^{-1}$~\cite{Panaitescu:2004sn}.
Then, the deceleration radius for the fast velocity ejecta is estimated to be $R_{\rm dec} \simeq 2.1 \times 10^{18} \mathcal{E}_{\rm fej, 51.5} A_{\star, 0.7}^{-1} \beta_{\rm fej, -0.5}^{-2} \rm~cm$. 
Here, we give a rough approximation for the magnetic field strength as $B_{\rm rs} \approx (\epsilon_{\rm B} 4 \pi \varrho_{\rm cbm} \beta_{\rm fs}^2 c^2)^{1/2}$, where we have used the energy density balance, i.e., $\varrho_{\rm fej} \beta_{\rm rs}^2=\varrho_{\rm cbm} \beta_{\rm fs}^2$. 
Note that the reverse shock velocity is measured in the ejecta rest frame. 
%where we assume the density of the shocked medium $\varrho_{\rm rs} \simeq \varrho_{\rm cbm}(r_{\rm dec})$ is equal to the density of CBM at the deceleration radius and the mean velocity of the shocked medium can be estimated from the ejecta velocity as $\beta_{\rm rs} \simeq \beta_{\rm ej}$.
%The characteristic Lorentz factor of the shock-accelerated electrons is
%\begin{eqnarray}
%\gamma_m &=& \frac{\epsilon_e}{f_e} \frac{p-2}{p-1} \frac{m_p}{m_e} \beta_{\rm rs}^2 \nonumber \\ &\simeq & 131 \epsilon_{e, -1} f_{e, -1}^{-1} \beta_{\rm rs, -0.3}^2,
%\end{eqnarray}
%where $p = 2.4$ is the power-law index of the accelerated electrons. 
Then we can derive the characteristic synchrotron frequency in the engine frame,
\begin{eqnarray}
\nu_m &\simeq& 1.5 \times 10^{10} \mathcal{E}_{\rm fej, 51.5}^{-1} \epsilon_{e, -1}^{2} f_{e, -2}^{-2} \epsilon_{\rm B, -0.3}^{1/2} A_{\star, 0.7}^{3/2} \nonumber \\ &\times& \beta_{\rm rs, -0.5}^4 \beta_{\rm fej, -0.5}^3 \rm~Hz,
\end{eqnarray}
where we assume the accelerated electrons following a power-law distribution with spectral index $s = 2.4$ and $\gamma_m = ((s-2)/(s-1))(\epsilon_e / f_e) (m_p / m_e) \beta_{\rm rs}^2$.
Similarly, the electron cooling frequency is
\begin{eqnarray}
\nu_c &\simeq& 1.1 \times 10^{14} \mathcal{E}_{\rm fej, 51.5} \epsilon_{\rm B, -0.3}^{-3/2} A_{\star, 0.7}^{-5/2} \nonumber \\ &\times& \beta_{\rm fej, -0.5}^{-5} \rm~Hz.
\end{eqnarray}
and the synchrotron self-absorption frequency is 
\begin{eqnarray}
\nu_a &\simeq& 2.5 \times 10^7 \mathcal{E}_{\rm fej, 51.5}^{-1} \varepsilon_{e,-1}^{7/5} f_{e, -2}^{-7/5} \epsilon_{B,-0.3}^{1/5}  \nonumber \\ &\times& A_{\star, 0.7}^{9/5} \beta_{\rm rs, -0.5}^{17/5} \beta_{\rm fej, -0.5}^{18/5}\rm~Hz.
\end{eqnarray}
The radiation luminosity per unit energy at characteristic frequency can be estimated to be
\begin{eqnarray}
L_{\varepsilon, \rm max} \simeq 1.2 \times 10^{56} f_{e, -2} \epsilon_{\rm B, -0.3}^{1/2} A_{\star, 0.7}^{3/2} \beta_{\rm fej, -0.5} \rm~s^{-1}.
\end{eqnarray}
We show the differential photon number density in Fig.~\ref{fig:RS_photon_density} which is estimated using following parameter sets, 

TRSN-A: $\mathcal{E}_{\rm fej} = 3 \times 10^{51} \rm~erg$, $\beta_{\rm fej} = 0.8$, $A_\star = 1$, $\epsilon_e = 0.1$, $f_e = 0.01$, $\epsilon_B = 0.1$ and $s = 2.4$.

TRSN-B: $\mathcal{E}_{\rm fej} = 3 \times 10^{51} \rm~erg$, $\beta_{\rm fej} = 0.3$, $A_\star = 5$, $\epsilon_e = 0.1$, $f_e = 0.01$, $\epsilon_B = 0.5$ and $s = 2.4$.

\section{\label{sec:three}UHECR nuclei from engine-driven SNe}
\subsection{\label{sec:three_A}Acceleration and survival of UHECR nuclei}
Charged nuclei can be accelerated to higher energies through the first-order Fermi acceleration mechanism~\cite{Bell:1978zc, Blandford:1987pw, Waxman:1998yy}.
The maximum acceleration energy is determined by the condition that the acceleration time scale should be smaller than dynamical time scale $t_{\rm acc} \leq t_{\rm dyn}$, as well as various energy cooling time scales, $t_{\rm acc} \leq t_{\rm cool} \equiv 1 / (t_{\rm ad}^{-1} + t_{A\gamma}^{-1} + t_{\rm syn}^{-1} + t_{\rm IC}^{-1})$.
Here $t_{\rm acc}$ is the acceleration time, $t_{\rm dyn}$ is the dynamical time, $t_{\rm cool}$ is the total energy cooling time, $t_{\rm ad} \sim t_{\rm dyn}$ is the adiabatic cooling time, $t_{\rm syn}$ is the synchrotron cooling time, $t_{\rm IC}$ is the inverse Compton cooling time, and $t_{A\gamma}$ is the photohadronic cooling time~\cite{Zhang:2017hom, Murase:2007yt}.
In the case of the age-limited acceleration ($t_{\rm cool} \geq t_{\rm dyn}$), the engine frame maximum energy of nuclei accelerated in jets can be estimated as,
\begin{eqnarray}
E_{A, \rm max}^{\rm jet} &\simeq& \eta^{-1} Z e B_\times (R_\times / \Gamma_\times) \Gamma_\times \nonumber \\ &\simeq& 7 \times 10^{20} \eta^{-1} (Z/26) \epsilon_{B,-1.3}^{1/2} \nonumber \\ &\times& \mathcal{E}_{{\rm k}, 51.5}^{1/4} \varrho_{\rm cbm, 1}^{1/4} T_{4}^{1/4} \rm~eV,
\end{eqnarray}
where $\eta \geq 1$ is the acceleration efficiency~\cite{Murase:2005hy}, $Z$ is the nucleus charge number, $R_\times / \Gamma_\times$ is the comoving width of the shocked ejecta. While the maximum energy of nuclei accelerated in transrelativistic ejecta is estimated to be,
\begin{eqnarray}
E_{A, \rm~max}^{\rm fej} &\simeq& \eta^{-1} Z e B_{\rm rs} R_{\rm dec} \beta_{\rm fej} \nonumber \\ &\simeq& 8.3 \times 10^{19} \eta^{-1} (Z/26) \epsilon_{\rm B, -0.3}^{1/2} \nonumber \\ &\times& A_{\star, 0.7}^{1/2}
 \beta_{\rm fej, -0.5}^2 \rm~eV.
\end{eqnarray}
We see that the acceleration energy of heavy nuclei can exceed $\sim 10^{20}\rm~eV$ in both jets and transrelativistic ejecta. Note that the maximum energy is sensitive to the ejecta velocity, $E_{A, \rm~max}^{\rm fej} \propto \beta_{\rm fej}^2$ in the latter case, and the acceleration of UHECR nuclei to ultrahigh energies is much difficult for lower-velocity ejecta.
Note that most of the ejecta kinetic energy is taken by ejecta with relatively lower velocities, which may make the situation even worse~\cite{Wang:2007ya}. Considering the diversity of engine-driven SNe and the uncertainty of the ejecta profile~\cite{Lazzati:2011ay}, we simply assume that high-velocity ejecta ($\beta_{\rm ej} \geq 0.3$) takes a significant fraction of the total outflow energy. The situation can also be alleviated by denser CBM and/or stronger magnetic field amplification.

Another requirement is that the energy budget of UHECR nuclei from engine-driven SNe should explain the observation results. The total energy of CR nuclei per event is $\mathcal{E}_{\rm CRacc} = 6 \times 10^{50} \xi_{\rm CRacc, -0.7} \mathcal{E}_{\rm k, 51.5} \rm~erg$ where $\mathcal{E}_{\rm k} = 3 \times 10^{51}\rm~erg$ is the kinetic energy of the jet or transrelativistic ejecta and $\xi_{\rm CRacc} = 1/5$ is the energy fraction of accelerated CRs. 
Then, we can derive the energy injection rate density, $Q_{\rm UHECR} = \mathcal{E}_{\rm CResc} \dot{\rho} \simeq 2 \times 10^{44} \mathcal{E}_{\rm CResc, -1} \dot{\rho}_{3.5} \rm~erg~Mpc^{-3}~yr^{-1}$, where $\dot{\rho}$ is the local event rate of engine-drivne SNe. We found the energy injection rate ensity $Q_{\rm UHECR}$ is comparable to the required energy budget of UHECR nculei~\cite{Murase:2008sa,Katz:2008xx, Zhang:2017moz}.

In Fig.~\ref{fig:source_spectrum_cr}, we show the energy spectrum of UHECR nuclei from engine-driven SNe. The escaping cosmic rays show a very hard energy spectrum if we adopt the escape-limited acceleration scenario under the condition $t_{\rm esc} \leq t_{\rm dyn}$. We can see that cosmic rays around maximum energy $E_{A, \rm max}^{\rm (esc)} \simeq x_{\rm esc} E_{A, \rm max}^{\rm (acc)} \simeq 10^{18.3} x_{\rm esc, -0.2} \rm~eV$ can escape efficiently, where $x_{\rm esc}$ is the ratio of the escape boundary to the width of the shocked region~\cite{Ohira:2009rd, Zhang:2017moz}. The maximum escape energy adopted in this work can be achieved assuming the acceleration efficiency is $\eta = 8$ for the jet and $\eta=1$ for the transrelativistic ejecta.

\begin{figure}
\includegraphics[width=\linewidth]{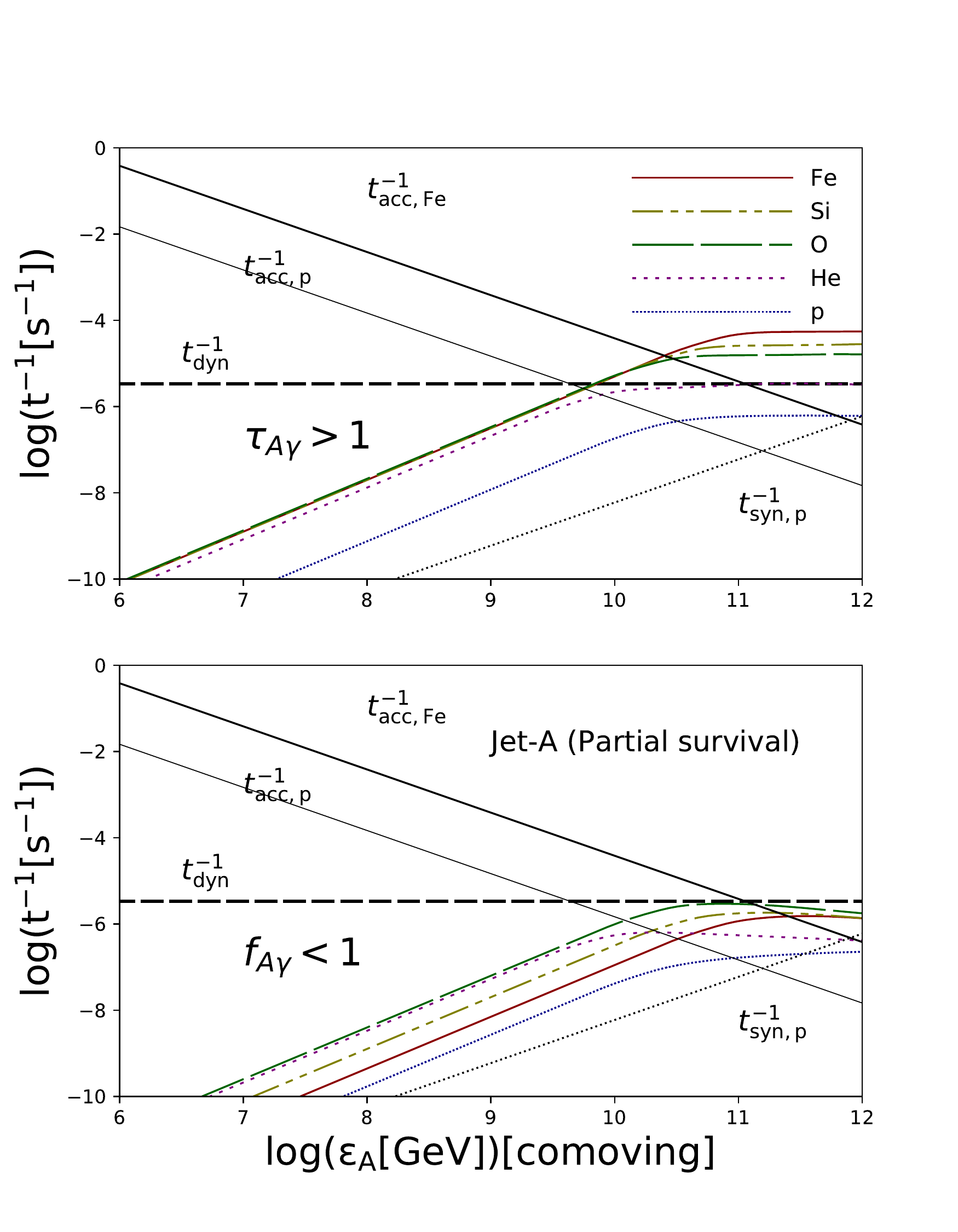}
\caption{Various time scales for five typical chemical species (Fe, Si, O, He, and proton) in the RS as a function of particle energy (measured in the engine frame) calculated from model Jet-A. We show the interaction time scales in the upper panel and energy loss time scales in the lower panel. The thin (thick) black line represents the acceleration time scale for proton (Fe). We show the synchrotron cooling time scale for proton as the dotted black line. Note that UHECR nuclei are in the partial survival regime, $\tau_{A\gamma} >1$ and $f_{A\gamma} < 1$, where a fraction of nuclei can survive. \label{fig:Jet-A}}
\end{figure}

\begin{figure}
\includegraphics[width=\linewidth]{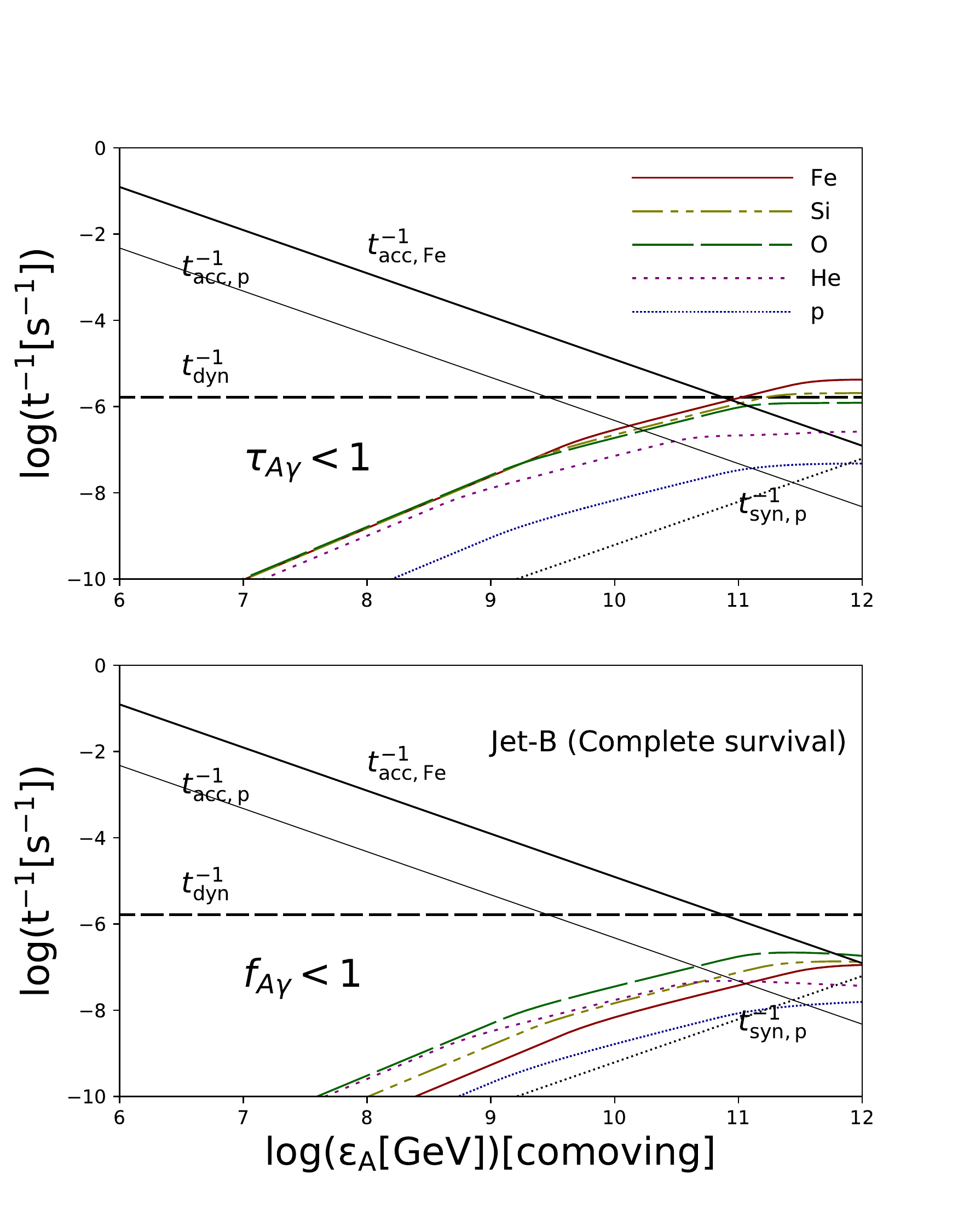}
\caption{Same as Fig.~\ref{fig:Jet-A}, but calculated from model Jet-B. Note that UHECR nuclei are in the complete survival regime, $\tau_{A\gamma} <1$ and $f_{A\gamma} < 1$, where nearly all UHECR nuclei can survive. \label{fig:Jet-B}}
\end{figure}

\begin{figure}
\includegraphics[width=\linewidth]{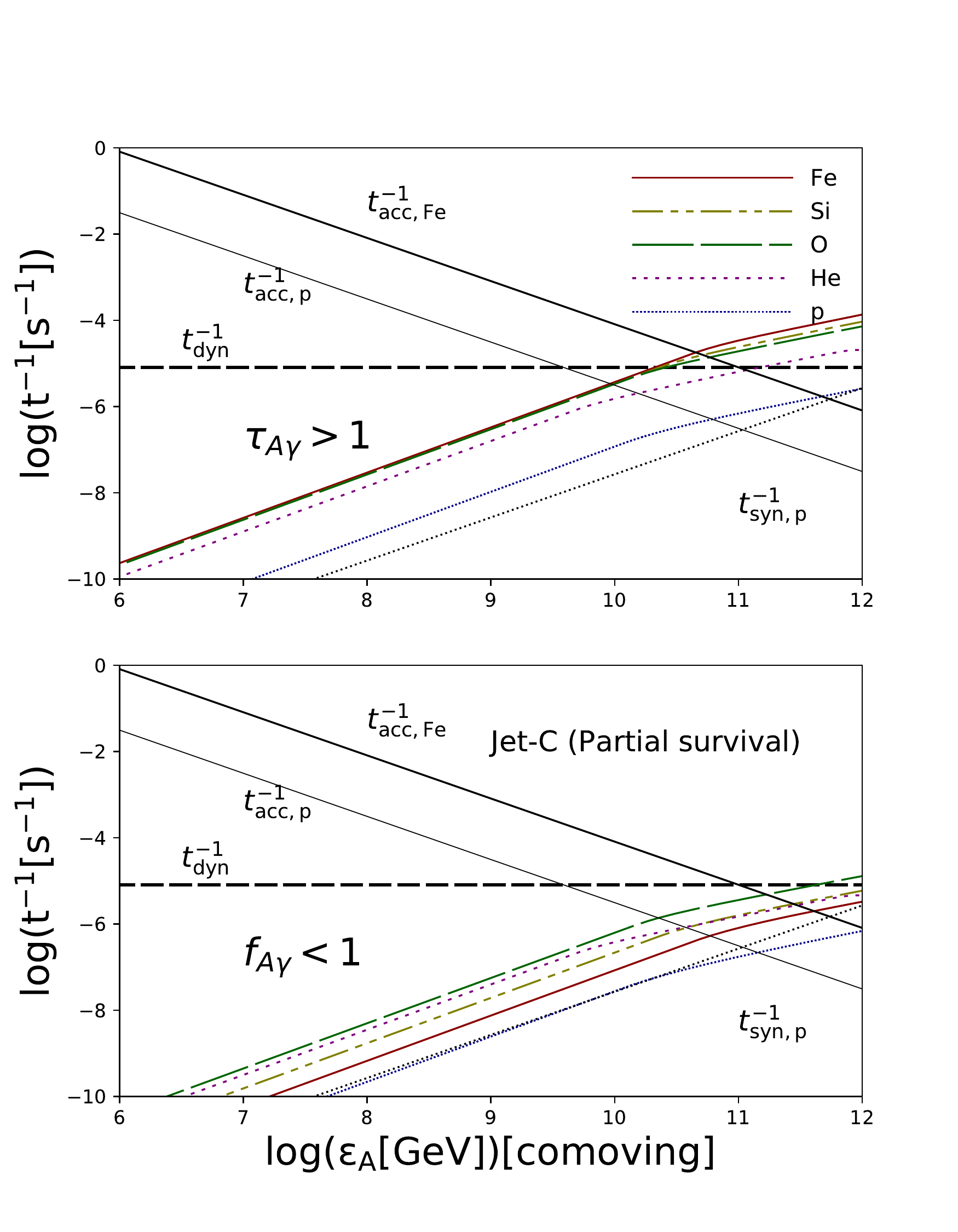}
\caption{Same as Fig.~\ref{fig:Jet-A}, but calculated from model Jet-C. Note that UHECR nuclei are in the partial survival regime, $\tau_{A\gamma} >1$ and $f_{A\gamma} < 1$, where a fraction of nuclei can survive. \label{fig:Jet-C}}
\end{figure}

The dominant energy cooling process of UHECR nuclei is photodisintegration. Using the giant dipole resonance (GDR) approximation, the optical depth of UHECR nuclei which is defined as $\tau_{A\gamma} \equiv  t_{\rm dyn} / t_{A\gamma-{\rm int}}$ can be estimated analytically~\cite{Murase:2010gj}
\begin{eqnarray}
\tau_{A\gamma} &\approx& \frac{2 (R_\times / \Gamma_\times) n_0 \varepsilon_0}{ (1 + \alpha)} \sigma_{\rm GDR} \frac{\Delta \bar{\varepsilon}_{\rm GDR}}{\bar{\varepsilon}_{\rm GDR}} \left(\frac{E_A}{E_{A0}}\right)^{\alpha - 1} \nonumber \\ 
&\simeq& 4.6~\mathcal{E}_{{\rm k}, 51.5}^{1/4} \varrho_{\rm cbm, 1}^{-1/4} T_{4}^{1/4} \nonumber \\ &\times& \left(\varepsilon_0 / 10^{-2}\rm~eV \right) \left(E_A / 10^{20}\rm~eV \right)^{-1/3},
\end{eqnarray}
where we adopt model Jet-A and Oxygen nuclei with $\sigma_{\rm GDR} = 3.1 \times 10^{-27} \rm~cm^2$, $\Delta \bar{\varepsilon}_{\rm GDR} = 6 \rm~MeV$, and $\bar{\varepsilon}_{\rm GDR} = 22.35\rm~MeV$~\cite{Batista:2016yrx}. 
The target photon number density is expressed as $dn/d\varepsilon = n_0 (\varepsilon / \varepsilon_0)^{-\alpha}$, where $\alpha = 2 / 3$ is the spectral index. The effective optical depth is defined as $f_{A\gamma} \equiv  t_{\rm dyn} / t_{A\gamma}$, where $t_{A\gamma}$ is the energy loss time scale of UHECR nuclei. The relation between effective optical depth $f_{A\gamma}$ and optical depth $\tau_{A\gamma}$ can be expressed as~\cite{Murase:2010gj} 
\begin{eqnarray}
f_{A\gamma} &=& \tau_{A\gamma} \bar{\kappa}_{A\gamma} \nonumber \\ 
&\simeq& 0.9~\mathcal{E}_{{\rm k}, 51.5}^{1/4} \varrho_{\rm cbm, 1}^{-1/4} T_{4}^{1/4} \left(\varepsilon_0 / 10^{-2}\rm~eV \right) \nonumber \\ &\times&   \left(E_A / 10^{20}\rm~eV \right)^{-1/3} \left( \bar{\kappa}_{A\gamma} / 0.2 \right),
\end{eqnarray}
where $\bar{\kappa}_{A\gamma}$ is the inelasticity which represents average energy losses of the photodisintegration process~\cite{Zhang:2017hom}.
We can see that UHECR nuclei are in the partial survival regime for model Jet-A which satisfy the condition $\tau_{A\gamma} >1$ and $f_{A\gamma} < 1$~\cite{Zhang:2017hom, Zhang:2017moz} (see Fig.~\ref{fig:Jet-A}). 
The complete survival of UHECR nuclei is possible when $\tau_{A\gamma} < 1$ for model Jet-B (see Fig.~\ref{fig:Jet-B}). 

\begin{figure}
\includegraphics[width=\linewidth]{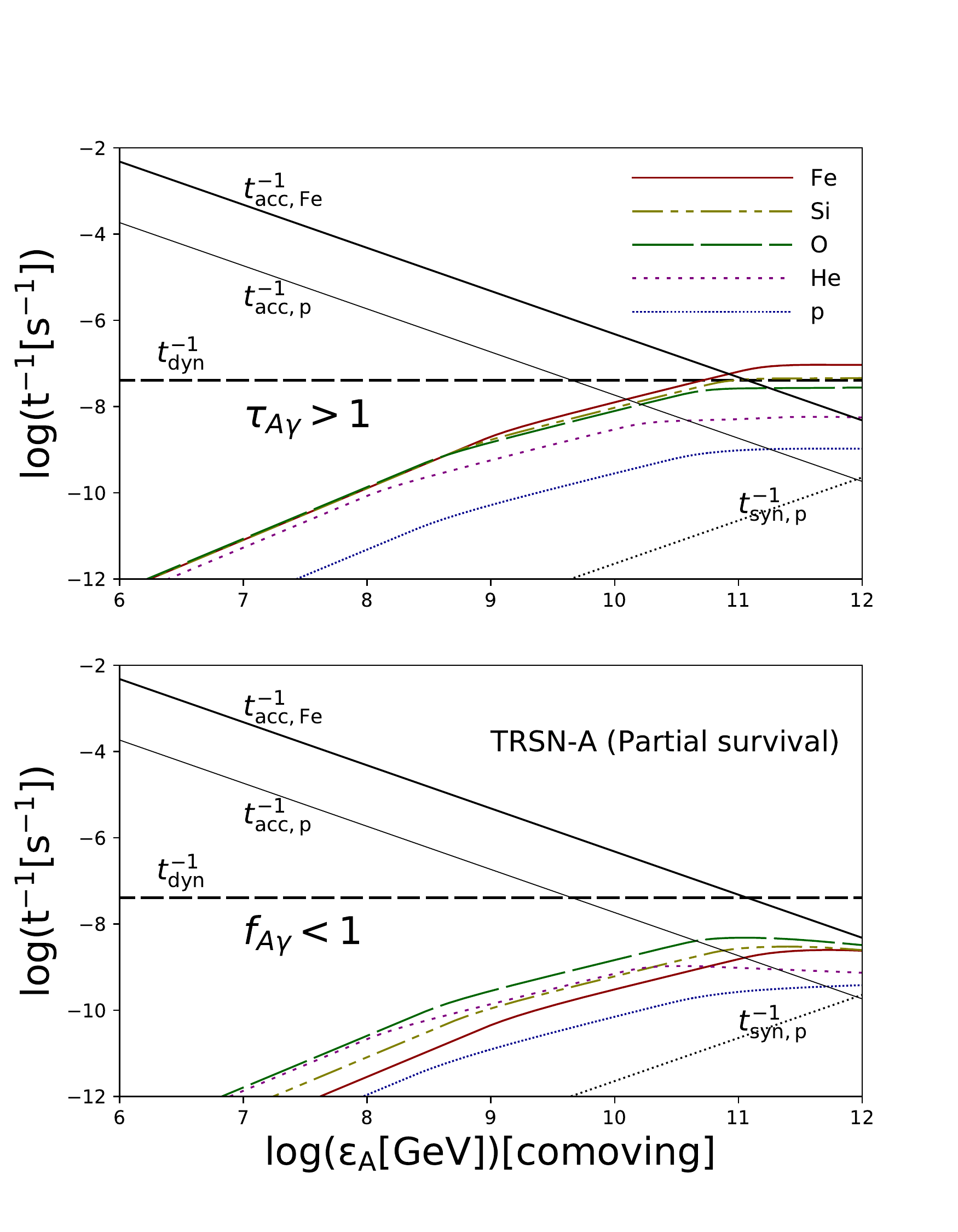}
\caption{Same as Fig.~\ref{fig:Jet-A}, but calculated from model TRSN-A. Note that UHECR nuclei are in the partial survival regime, where a fraction of UHECR nuclei can survive. \label{fig:TRSN-A}}
\end{figure}

\begin{figure}
\includegraphics[width=\linewidth]{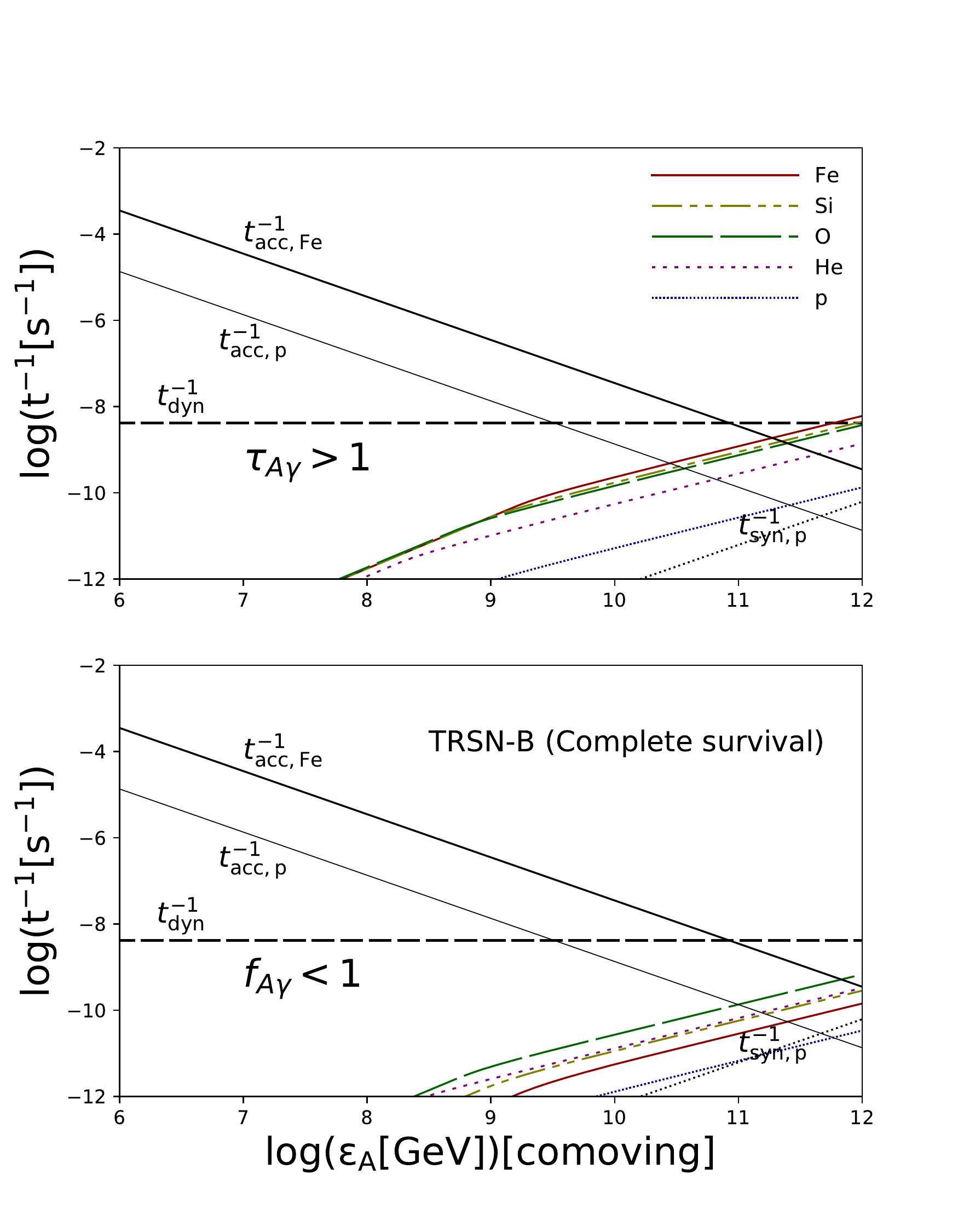}
\caption{Same as Fig.~\ref{fig:Jet-A}, but calculated from model TRSN-B. Note that UHECR nuclei are in the complete survival regime, where nearly all nuclei can survive. \label{fig:TRSN-B}}
\end{figure}

We also show the survival of UHECR nuclei for model TRSN-A (partial survival regime) in Fig.~\ref{fig:TRSN-A} and TRSN-B (complete survival regime) in Fig.~\ref{fig:TRSN-B}, respectively. We can see that the survival of UHECR nuclei is possible in both jets and transrelativistic ejecta for various parameter sets. One limitation of our study is that in our calculation we assume an ISM-like density profile of CBM for jets and a wind-like density profile for transrelativistic ejecta. Details of the survival are affected by the ambient density.  For jets, the wind-like density profile is dangerous for the conventional value $A_\star =1$ because the density becomes too high.  
Our results are not sensitive to the specific value of electron spectral index, even though it may slightly affect the electron injection frequency.

\subsection{\label{sec:three_B}Comparison with Auger and TA data}
UHECRs have been observed on Earth by various experiments~\cite{Kotera:2011cp, Anchordoqui:2018qom}, such as the Pierre Auger Observatory (Auger)~\cite{ThePierreAuger:2015rma} and Telescope Array (TA) collaboration~\cite{AbuZayyad:2012kk}, which are the two largest cosmic ray detectors at present. 
It has been shown that the energy spectrum of UHECR nuclei becomes flatter at energy around ``ankle'' $\sim 4 \times 10^{18}\rm~eV$, with a suppression at $\sim 5 \times 10^{19}\rm~eV$. 
However, there may be a significant statistical discrepancy in the energy spectra between Auger and TA around the suppression energy. The TA results show a relatively higher flux, and the reason is still unclear yet~\cite{TheTelescopeArray:2018dje}.

In this section, we aim to fit the energy spectrum and composition of UHECR nuclei detected by Auger~\cite{Aab:2017njo} and TA~\cite{Fukushima:2015bza, Abbasi:2018nun}.
For TA data, we decrease the flux of UHECR nuclei by a fraction of $\sim 13\%$ in order to be compatible with the Auger data around ``ankle'' region.
Even though it is difficult to make a direct comparison of the mean and standard deviation of the measured $X_{\rm max}$ of Auger~\cite{Aab:2017njo} and TA~\cite{Abbasi:2018nun} directly, which may be affected by the particular treatment of biases and detector efficiencies~\cite{TheTelescopeArray:2018dje}, we show the results derived by Auger and TA together for demonstrative purposes.

In this work, we propagate UHECR nuclei using the state-of-the-art Monte Carlo code CRPropa 3~\cite{Batista:2016yrx}.
For simplicity, we consider $1\rm D$ propagation, where the sources are uniformly distributed in the universe at each redshift bin and neglect the effects of large-scale magnetic fields. We adopt a semi-analytic extragalactic background light (EBL) model developed by Ref.~\cite{Gilmore2012}. This model is based on the hierarchical structure formation scenario, taking into account galaxy formation and evolution. 
We performed $\chi^2$ analysis as in Ref.~\cite{Zhang:2017moz, Fang:2017zjf} to show the quality of our fitting. The spectrum energy fitting range is from $10^{18.45}~\rm eV$ to $10^{21.15}~\rm eV$, where the uncertainties on the flux are taken from the measurements of Auger~\cite{Aab:2017njo} and TA~\cite{Fukushima:2015bza}. As in Ref.~\cite{Heinze:2015hhp}, we take into account the effect of systematic uncertainty on the energy scale, $E = E^{\rm Auger} (1 + \delta_E)$. 
We derived the first two moments of the $X_{\rm max}$ distributions using the EPOS-LHC hadronic interaction model~\cite{DeDomenico:2013wwa, Zhang:2017hom}, and compared to the measurements from Auger~\cite{Zhang:2017moz} and TA~\cite{Abbasi:2018nun}, respectively.

\begin{figure}
\includegraphics[width=\linewidth]{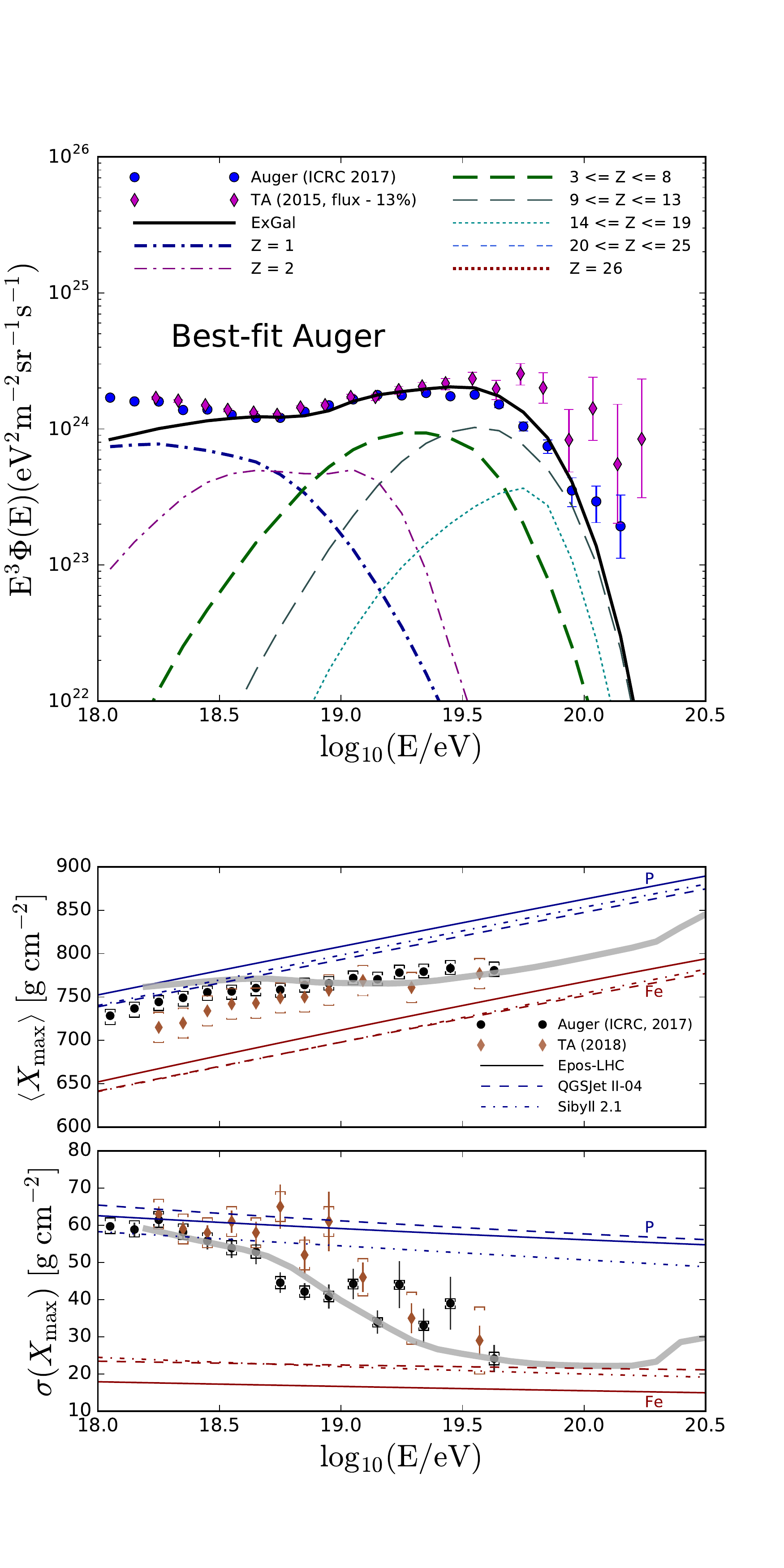}
\caption{Upper panel: the energy spectrum of UHECR nuclei from model Jet-B, where UHECR nuclei are in the complete survival regime.
Lower panel: the mean and standard deviation of the measured $X_{\rm max}$ distributions as a function of energy. We use the measurements from Auger (blue and black dots)~\cite{Aab:2017njo} and TA (magenta and sienna dimonds)~\cite{Fukushima:2015bza, Abbasi:2018nun}.
 \label{fig:Jet-B_spectrum}}
\end{figure}

\begin{figure}
\includegraphics[width=\linewidth]{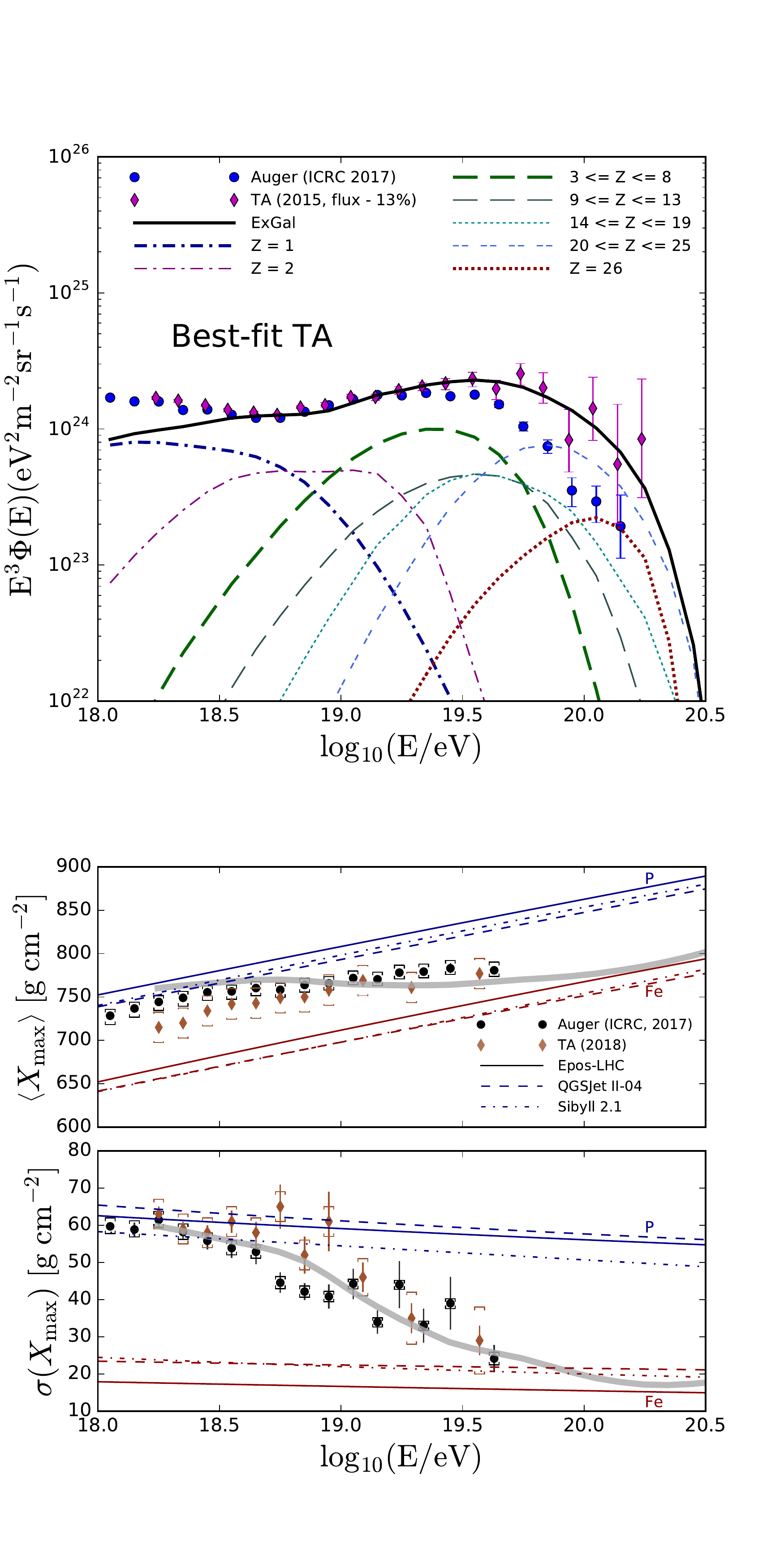}
\caption{Same as Fig.~\ref{fig:Jet-B_spectrum}, but for model TRSN-B, where UHECR nuclei are in the complete survival regime. 
\label{fig:TRSN-B_spectrum}}
\end{figure}

The results are shown in Fig.~\ref{fig:Jet-B_spectrum}-\ref{fig:TRSN-B_spectrum}, where we consider the complete survive regime, Jet-B and TRSN-B.
The change to a lighter composition in the $X_{\rm max}$ distributions is caused by the photodisintegration process of nuclei from nearest sources, as explained in Ref.~\cite{Zhang:2017moz}.
We find model Jet-B can fit the Auger data very well, $\chi^2 / \rm d.o.f. = 1.5$, where the best-fitting parameters is $E_{p, \rm max}^{\rm esc} = 18.3$ and $\delta_E = 0.13$, as shown in Fig.~\ref{fig:Jet-B_spectrum}.
On the other hand, model TRSN-B gives the best fit to TA data, $\chi^2 / \rm d.o.f. = 0.9$, where the best-fitting parameters are $E_{p, \rm max}^{\rm esc} = 18.3$ and $\delta_E = 0.01$, see Fig.~\ref{fig:TRSN-B_spectrum}. The difference between Jet-B and TRSN-B is the composition, in which the latter has a significant fraction of iron-group nuclei. 
Compared to intermediate mass group nuclei, iron-group nuclei can be accelerated to higher energies, with longer energy loss lengths. 
Although the discrepancy in the energy spectra between the Auger and TA data is still at the $\sim3\sigma$ level, our results imply that it could in principle be caused by more engine-driven SNe (with more iron nuclei) in the northern sky within the local universe~\cite{Globus:2016gvy}. 
The injection of heavier nuclei depend on the angular momentum of the progenitor core. Canonical GRB preferentially occur in low-metallicity environments, but LL GRBs and engine-driven SNe may largely occur in starburst galaxies that are often metal polluted.   

Note that both Jet-B and TRSN-B models essentially correspond to propagation-only models in Ref.~\cite{Boncioli:2018lrv}. In this work, we also consider the partial survival regime, where only a fraction of nuclei can survive and the energy spectrum of escaping UHECR nuclei may be affected by the nuclear cascade effect. However, even for Jet-A and TRSN-A models, it turns out that the final energy spectrum of UHECR nuclei is not too much affected by the nuclear cascade process if we adopt the parameters listed in the end of Sec.~\ref{sec:two_A} and Sec.~\ref{sec:two_B}.

\section{\label{sec:four}Neutrinos from engine-driven SNe}
The total cross section of the photomeson production process can be decomposed into three parts: $\Delta$-resonance region at peak energy $\bar{\varepsilon}_\Delta \simeq 340 \rm~MeV$ (with a contribution from direct production), higher resonances with two or three peaks in the energy range $500 {\rm~MeV} < \bar{\varepsilon} < 1.5 {\rm~GeV}$, multipion production region at much higher energies ($> 1.5 {\rm~GeV}$) (see, e.g., Ref.~\cite{Rachen1996} for details). 
The box approximation is commonly used, in which $\sigma_{p\gamma}  \approx \sigma_\Delta \delta (\bar{\varepsilon} - \bar{\varepsilon}_\Delta) \Delta \bar{\varepsilon}_\Delta$. Here $\sigma_\Delta \sim 4.4 \times 10^{28} {\rm~ cm^2}$, $\bar{\varepsilon}_\Delta \sim 0.34 {\rm~GeV}$, and $\Delta \bar{\varepsilon}_\Delta \sim 0.2 {\rm~GeV}$~\cite{Murase:2010gj}.
Note that the branching ratio between neutron pion $\pi^0$ production channel and charged pion $\pi^\pm$ production channel can effectively approach to $1:1:1$ owing to multipion production~\cite{Rachen1996,Murase:2005hy}. The final neutrino energy spectrum from charged pion and muon decay can be calculated numerically through the code SOPHIA~\cite{Mucke:1999yb} or GEANT4~\cite{Murase:2005hy}. 

For nuclei, we assume that the cross section on nuclei scales linearly with their mass, $\sigma_{\rm A\gamma} (\bar{\varepsilon}) = A \sigma_{\rm p\gamma} (\bar{\varepsilon})$ and the inelasticity is assumed to have $\kappa_{\rm A\gamma} (\bar{\varepsilon}) = (1/A)  \kappa_{\rm p\gamma} (\bar{\varepsilon})$. Note that the cross section of the photomeson production on nuclei can be different from the simple scaling relations if we consider the more complicated Fermi motion of nucleons and other inmedium modifications~\cite{Rachen1996, Kampert:2012fi, Biehl:2018xjv}. 

Considering the photodisintegration of UHECR nuclei, the generated neutrinos can be divided into two parts, one is the direct neutrino contribution due to the photomeson production on nuclei, and another is the indirect neutrino contribution by secondary protons/neutrons (that are stripped off from their parent nuclei due to the photodisintegration process)~\cite{Murase:2010gj}.

In the first case, the all-flavor neutrino energy spectrum can be simply estimated using the following formula~\cite{Murase:2010gj},
\begin{equation}
\varepsilon_\nu^2 \frac{dN_\nu}{d\varepsilon_\nu} \approx \frac{3}{8} f_{\rm sup} f_{\rm mes} (\varepsilon_A) (1 - f_{A\gamma}(\varepsilon_A))\varepsilon_A^2 \frac{dN_A}{d\varepsilon_A},
\end{equation}
where $f_{\rm sup}$ is the suppression factor due to meson and muon cooling, $f_{\rm mes}(\varepsilon_A)$ is the effective optical depth for the photomeson production, and ($1- f_{A\gamma}$) represents the fraction of survival nuclei. 
The factor $3/8$ is due to the fact that half of the energy $\sim 1/2$ goes into charged pions and the pion decay convert $\sim 3/4$ of their energy into neutrinos. 
Assuming each nucleon has energy $\sim 1/A$ of the parent nucleus with $\varepsilon_A$, then the effective optical depth to the photomeson production on nuclei can be approximated by $f_{\rm mes} (\varepsilon_A) \approx f_{p\gamma} (\varepsilon_A / A)$. The typical neutrino energy is $\varepsilon_\nu \approx 0.05 \varepsilon_A / A$.
While in the second case, the all-flavor neutrino energy spectrum can be calculated as~\cite{Murase:2010gj}
\begin{equation}
\varepsilon_\nu^2 \frac{dN_\nu}{d\varepsilon_\nu} \approx \frac{3}{8}f_{\rm sup} f_{p\gamma}(\varepsilon_p) f_{A\gamma}(\varepsilon_A) \varepsilon_A^2 \frac{dN_A}{d\varepsilon_A},
\end{equation}
where $\varepsilon_p \approx \varepsilon_A / A$ and $f_{A\gamma}$ represents the fraction of photodisintegrated nuclei. In this case, the typical neutrino energy depends on the secondary proton energy $\varepsilon_\nu \approx 0.05 \varepsilon_p$. Note that the secondary protons/neutrons take a fraction $\sim f_{A\gamma} (\varepsilon_A)$ of the parent nuclei total energy.
The two cases can be combined into one formula
\begin{equation}
\varepsilon_\nu^2 \frac{dN_\nu}{d\varepsilon_\nu} \approx \frac{3}{8} f_{\rm sup} f_{p\gamma} (\varepsilon_A / A) \varepsilon_A^2 \frac{dN_A}{d\varepsilon_A}.
\end{equation}
under the approximation $f_{\rm mes} (\varepsilon_A) \simeq f_{p\gamma} (\varepsilon_A / A)$. 
The suppression factor accounts for the energy cooling on the primary pions and secondary muons, $f_{\rm sup} \simeq f_{\rm sup, \pi} f_{\rm sup, \mu}$~\cite{Kimura:2017kan}. The pion cooling suppression factor is calculated as $f_{\rm sup, \pi} = 1 - {\rm exp} (-t_{\pi, \rm cool} / t_{\pi, \rm decay})$, where $t_{\pi, \rm decay}$ is the decay time scale of pions. The mean life time for a pion to decay is $\tau_\pi = 2.6 \times 10^{-8} \rm~s$ in the pion rest frame, then the pion decay time scale in the shock comoving frame is $t_{\pi, \rm decay} = (\varepsilon_\pi / m_\pi) \tau_\pi$.  The dominant cooling process of pions are synchrotron cooling, $t_{\pi, \rm cool}^{-1} \sim t_{\pi, \rm syn}^{-1}$. The muon cooling suppression factor can be calculated using the same method.

We also perform numerical simulations in order to take into account the nuclear cascade process inside the source with the help of the publicly available Monte-Carlo code CRPropa~\cite{Kampert:2012fi, Batista:2016yrx}, where the photomeson production process is calculated using the SOPHIA package~\cite{Mucke:1999yb}. In CRPropa,  the mean free path of nuclei have the following scaling relations~\cite{Kampert:2012fi},
\begin{equation}
\lambda_{A, Z}^{-1} (\varepsilon_A) \simeq A \lambda_{p/n}^{-1}(\varepsilon_A/A),
\end{equation}
where $\lambda_p$, $\lambda_n$ are mean free paths for protons and neutrons, separately. The scaling relations equivalent to that there is only one of the nucleons in the parent nuclei actually interact with the target photons as we have declared before where the remained nuclei have mass $(A-1)$ and energy $\varepsilon_{A-1} = (A-1) / A \varepsilon_A$ assuming the conservation of the Lorentz factor.
For our purpose, we inject nuclei from the original point and propagate them following one direction until nuclei and the secondaries reach to the boundary of the acceleration site. 
In the numerical simulations, we avoid the synchrotron cooling on the intermediate products, such as pions and muons~\cite{Murase:2005hy}.

\subsection{Neutrino fluences}
The observed neutrino flucences can be estimated using the following formula~\cite{Murase:2007yt},
\begin{eqnarray}
E_\nu^2 {\phi}_{\nu} &=&  \frac{1+z}{4\pi d_L^2} \Gamma {\varepsilon^2_\nu} \frac{dN_\nu}{d\varepsilon_\nu},
\end{eqnarray}
where $d_L$ is the source luminosity distance, $\Gamma$ is the source Lorentz factor and ${\varepsilon^2_\nu} dN_\nu / d\varepsilon_\nu$ is the comoving frame neutrino energy spectrum inside the source. 

\begin{figure}
\includegraphics[width=\linewidth]{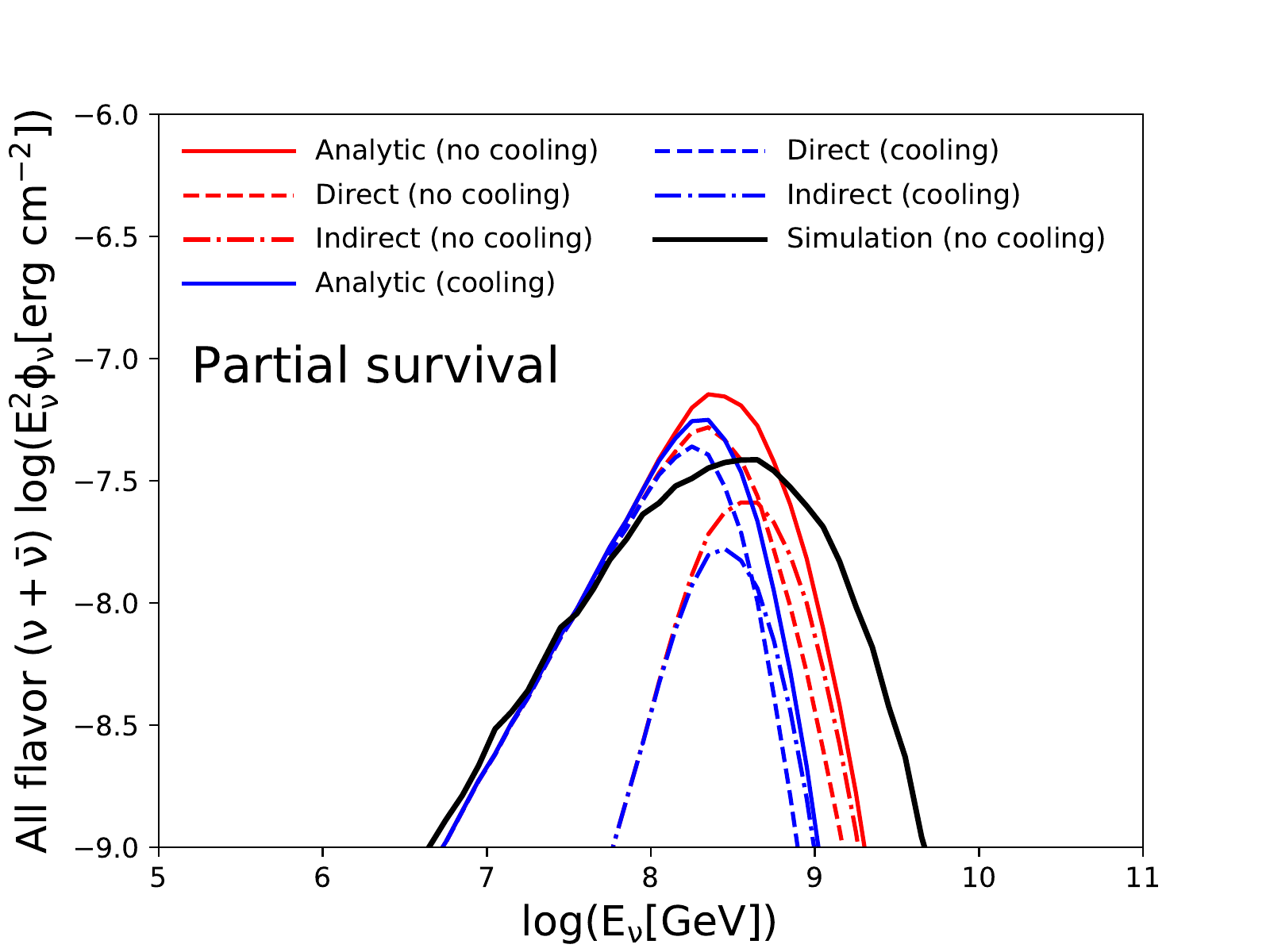}
\caption{All-flavor neutrino fluences estimated in the partial survival regime for model Jet-A. The red/blue line are the neutrino fluences calculated using analytic formula. The dashed red/blue lines represent the contributions from direct neutrinos, while the dot-dashed red/blue lines represent the contributions from indirect neutrinos. The black line are the neutrino fluences estimated using numerical simulations.
 \label{fig:neutrino_fluence_ModelA}}
\end{figure}

In Fig.~\ref{fig:neutrino_fluence_ModelA}, we show the results of all-flavor neutrino fluences observed at Earth (red line) emitted from one engine-driven SNe located at redshift $z = 0.005$. The neutrino fluence is normalized assuming the total energy of CR nuclei per event is $\mathcal{E}_{\rm CRacc} = 6 \times 10^{50} \xi_{\rm CRacc, -0.7} \mathcal{E}_{\rm k/fej, 51.5} \rm~erg$, as shown in Fig.~\ref{fig:source_spectrum_cr}.
 The partial survival regime corresponding to model Jet-A (or TRSN-A) is assumed. The effect of synchrotron cooling on intermediate pions and muons is shown (blue line). 
We can see that the synchrotron cooling is important in the partial survival regime which can suppress the neutrino fluences at highest energy range. The break energy due to pion/muon cooling can be estimated when $t_{\pi, \rm cool} = t_{\pi, \rm decay}$, $\varepsilon_{\nu, \pi} = 2.7 \times 10^{18} B_1^{-1} \rm~eV$ and $\varepsilon_{\nu, \mu} = 1.5 \times 10^{17} B_1^{-1} \rm~eV$, where $t_{\pi, \rm dec}$ is the decay time scale of pions or muons.
We also show the energy spectrum of direct neutrinos (dashed line) and indirect neutrinos (dot-dashed line), respectively. We find that in the regime of partial nucleus survival, the contribution from indirect neutrinos can have a dominant contribution to the observed neutrino fluences at higher energies~\cite{Murase:2010gj, Biehl:2017zlw}.

\begin{figure}
\includegraphics[width=\linewidth]{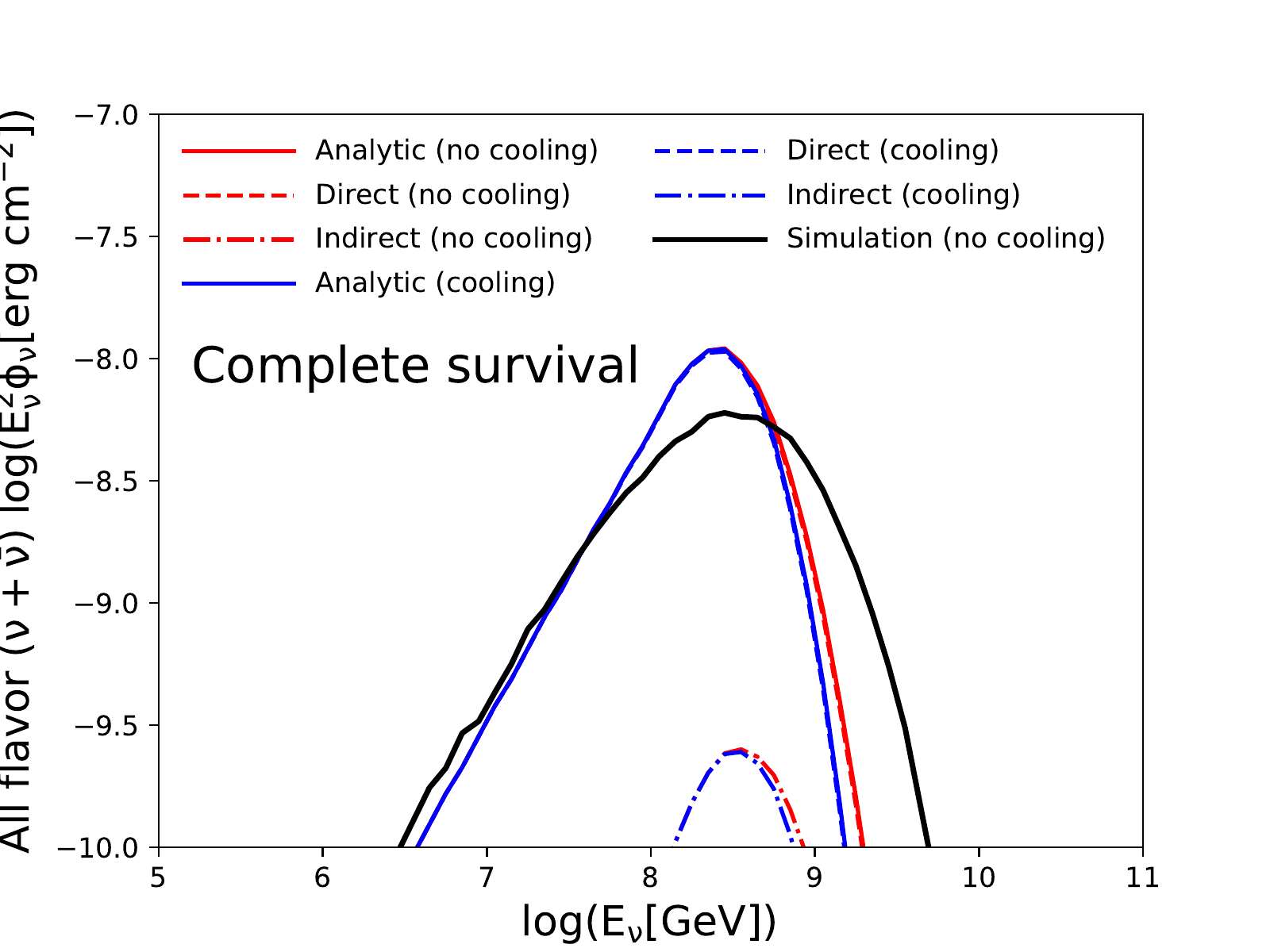}
\caption{Similar to Fig.~\ref{fig:neutrino_fluence_ModelA}, but estimated in the complete survival regime for model Jet-B.
 \label{fig:neutrino_fluence_ModelB}}
\end{figure}

Similar to Fig.~\ref{fig:neutrino_fluence_ModelA}, we show the all-flavor neutrino fluences estimated assuming complete survival regime corresponding to model Jet-B (or TRSN-B) in Fig.~\ref{fig:neutrino_fluence_ModelB}. We can see that the contribution to the neutrino fluences from indirect neutrinos can be neglected and the predicted neutrino fluences in the complete survival regime are about $\sim 1 - 2$ orders lower than in the partial survival regime.

We also present the neutrino fluences estimated through the numerical simulation which are indicated as black curve in Fig.~\ref{fig:neutrino_fluence_ModelA} and Fig.~\ref{fig:neutrino_fluence_ModelB}.
The neutrino fluences derived from numerical simulations which take into account the effect of nuclear cascade process are comparable to our analytic results. The difference at the higher energy range is due to the fact that the analytical approach does not take into account the contribution from multi-pion neutrino production as well as the energy spread of secondaries~\cite{Mucke:1999yb, AlvesBatista:2019rhs}. (The similar trend was found by Ref.~\cite{Murase:2007yt}.)
In the following, we will adopt the results calculated using our analytic formula which takes into account the effect of synchrotron cooling on intermediate pions and muons.

\begin{figure}
\includegraphics[width=\linewidth]{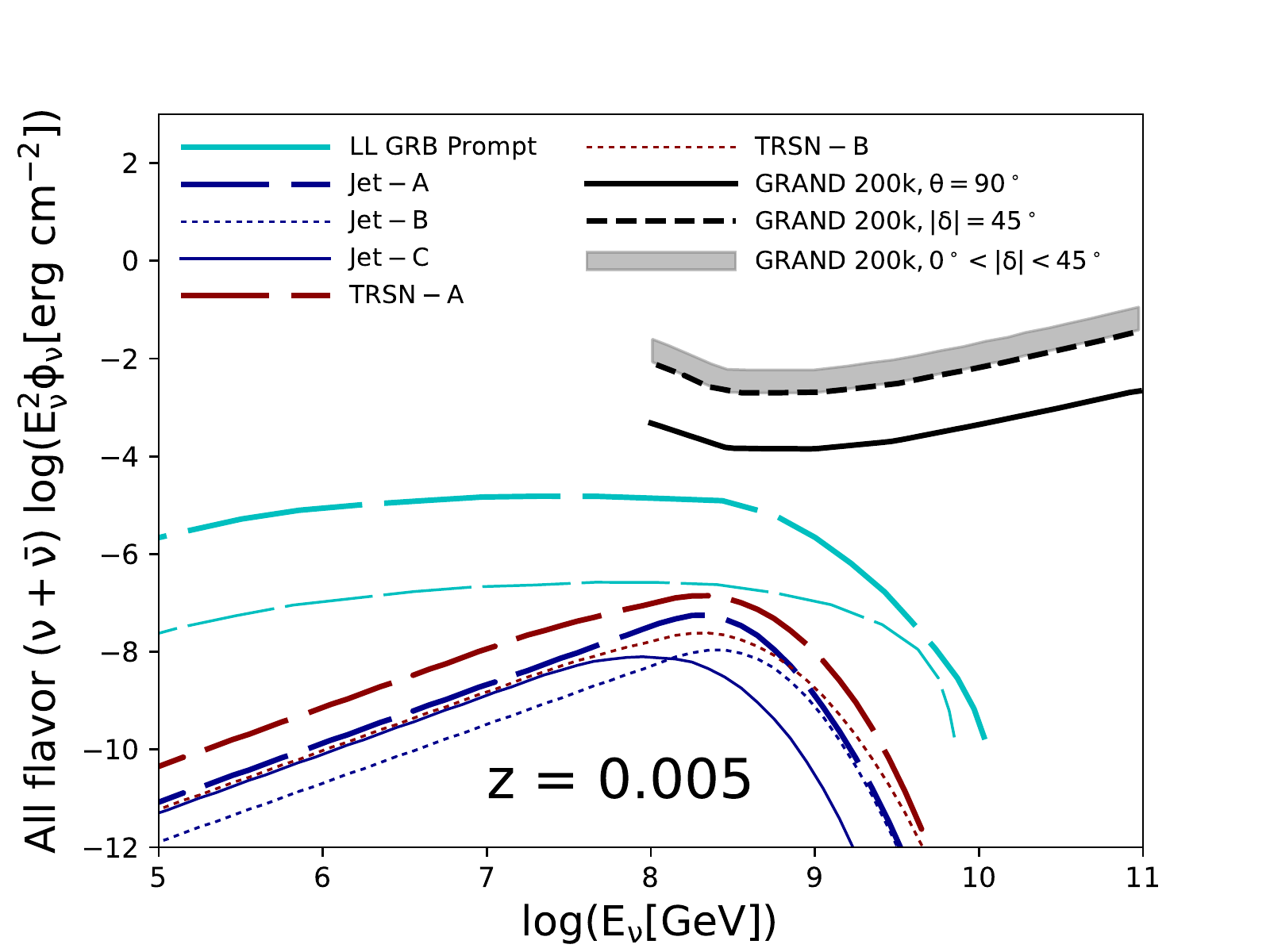}
\caption{All-flavor neutrino fluences from a single engine-driven SN (or LL GRB) located at redshitf $z = 0.005$. The thick and thin line represent the partial survival regime and complete survival regime, respectively. The cyan line is the results from Ref.~\cite{Murase:2008mr} for LL GRBs. The grey and black lines are the detection sensitivities of GRAND200k~\cite{Alvarez-Muniz:2018bhp}.\label{fig:pointsource_up}}
\end{figure}

In Fig.~\ref{fig:pointsource_up}, we show all-flavor neutrino fluences considering cooling effect derived in this work for one engine-driven SNe located at redshift $z = 0.005$ and the expected detection sensitivity from GRAND project~\cite{Alvarez-Muniz:2018bhp}.
We considered five models, Jet-A/B/C and TRSN-A/B, and we can see that the neutrino fluence depends on the composition and survival regime of UHECR nuclei.
We found that the neutrino fluence from engine-driven SNe is significantly lower than the detection sensitivity of GRAND, which makes it challenging to be detected as a point source of UHE neutrinos. 

\subsection{\label{sec:diff}Diffuse source neutrinos}
Considering the higher event rate of engine-driven SNe, it is reasonable to consider their contributions to the diffuse neutrino background, including both neutrinos produced inside sources (diffuse source neutrinos) and cosmogenic neutrinos.
Note that the normalization of the diffuse neutrinos is fixed under the condition that the escaped UHECR nuclei can reconcile with the observed energy spectrum and composition of UHECR nuclei measured by Auger simultaneously.
The flux of diffuse source neutrinos from engine-driven SNe can be estimated using the following formula,
\begin{eqnarray}
E_\nu^2 \Phi_\nu &=& \frac{1}{4\pi} \int_{z_{\rm min}}^{z_{\rm max}} dz  \frac{dV_c}{dz} \frac{F(z) \dot{\rho_0}}{1 + z} E_\nu^2 \phi_\nu,
\end{eqnarray}
where $E_\nu^2 \phi_\nu$ is the neutrino fluence derived in the previous section. The event rate of engine-driven SNe in the local universe is $\dot{\rho_0}$ and $F(z)$ is the redshift distribution parameter estimated from long GRBs~\cite{Sun:2015bda}. The redshift dependence of the comoving volume is
\begin{equation}
\frac{dV_c}{dz} = \frac{c}{H_0} \frac{4 \pi d_L^2}{(1+z)^2\sqrt{\Omega_M (1+z)^3 + \Omega_\Lambda}},
\end{equation}
where $d_L$ is the luminosity distance to the source.

\begin{figure}
\includegraphics[width=\linewidth]{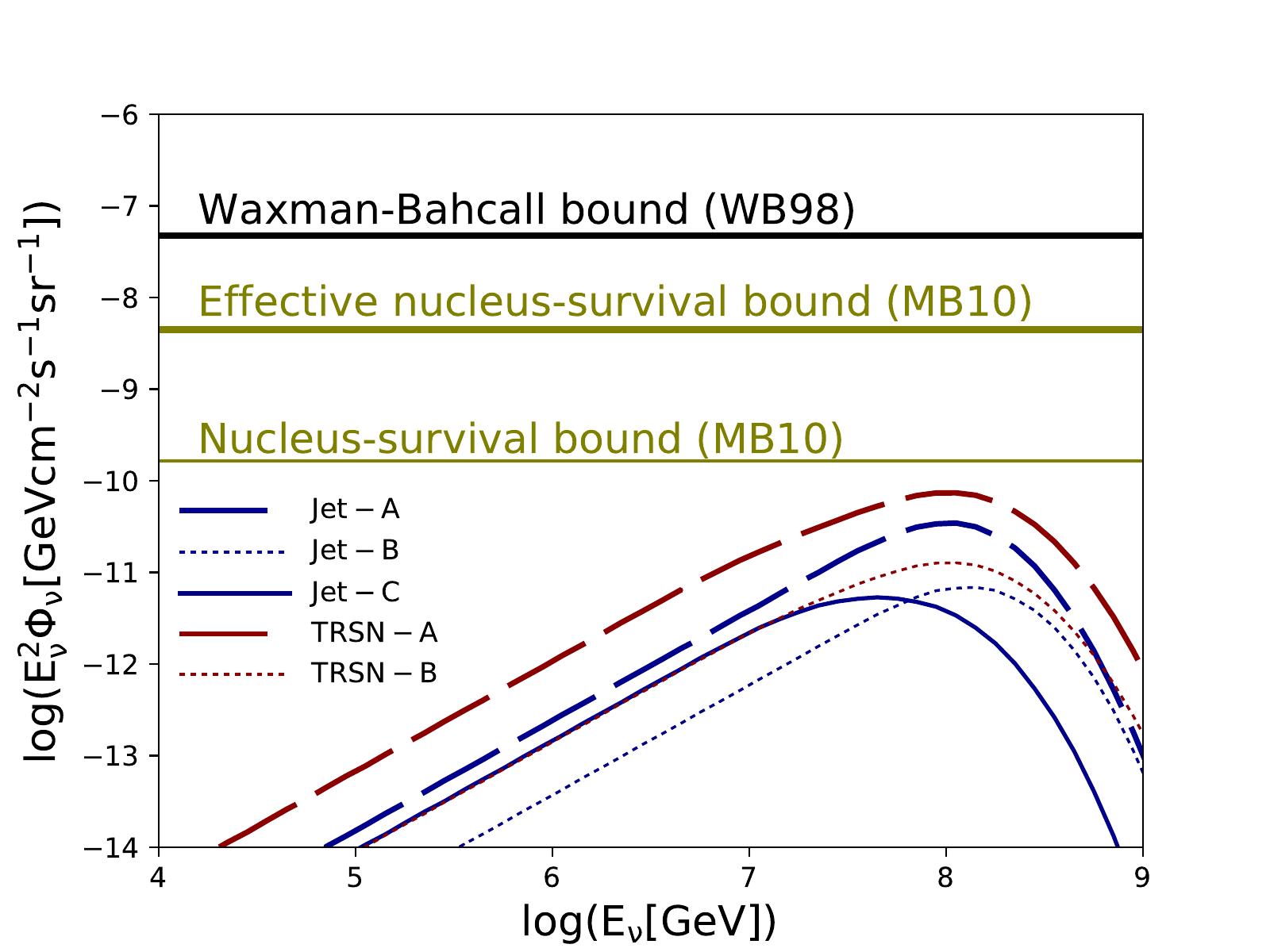}
\caption{Energy spectra of diffuse source neutrinos from engine-driven SNe or LL GRBs. The solid black line is the well-known Waxman-Bachall bound which is derived assuming pure proton composition (WB98)~\cite{Waxman:1998yy}. The solid olive line is the upper bound assuming UHECR nuclei have pure Silicion composition (MB10) as discussed in Ref.~\cite{Murase:2010gj} under the condition $f_{A\gamma} < 1$ (partial survival; thick olive line) and/or $\tau_{A\gamma} < 1$ (complete survival; thin olive line).
 \label{fig:diffuse_neutrino}}
\end{figure}

We show the energy spectrum of diffuse source neutrinos in Fig.~\ref{fig:diffuse_neutrino}.  The thick (thin) blue line represents diffuse neutrinos estimated from model Jet-A (Jet-B), while the thick (thin) red line represents diffuse neutrinos estimated from model TRSN-A (TRSN-B).
For comparison, we also show the well-known Waxman-Bahcall bound (black line) assuming fast-evolution scenario ($\xi_z \simeq 3$ in Ref.~\cite{Waxman:1998yy}) and photodisintegration bound (olive line) for pure Silicon composition~\cite{Murase:2010gj}. Note that in the non-evolution case, the diffuse neutrino fluxes can be $\sim 5$ times lower than in the fast-evolution case.

\subsection{\label{sec:cosmo}Cosmogenic neutrinos}
\begin{figure}
\includegraphics[width=\linewidth]{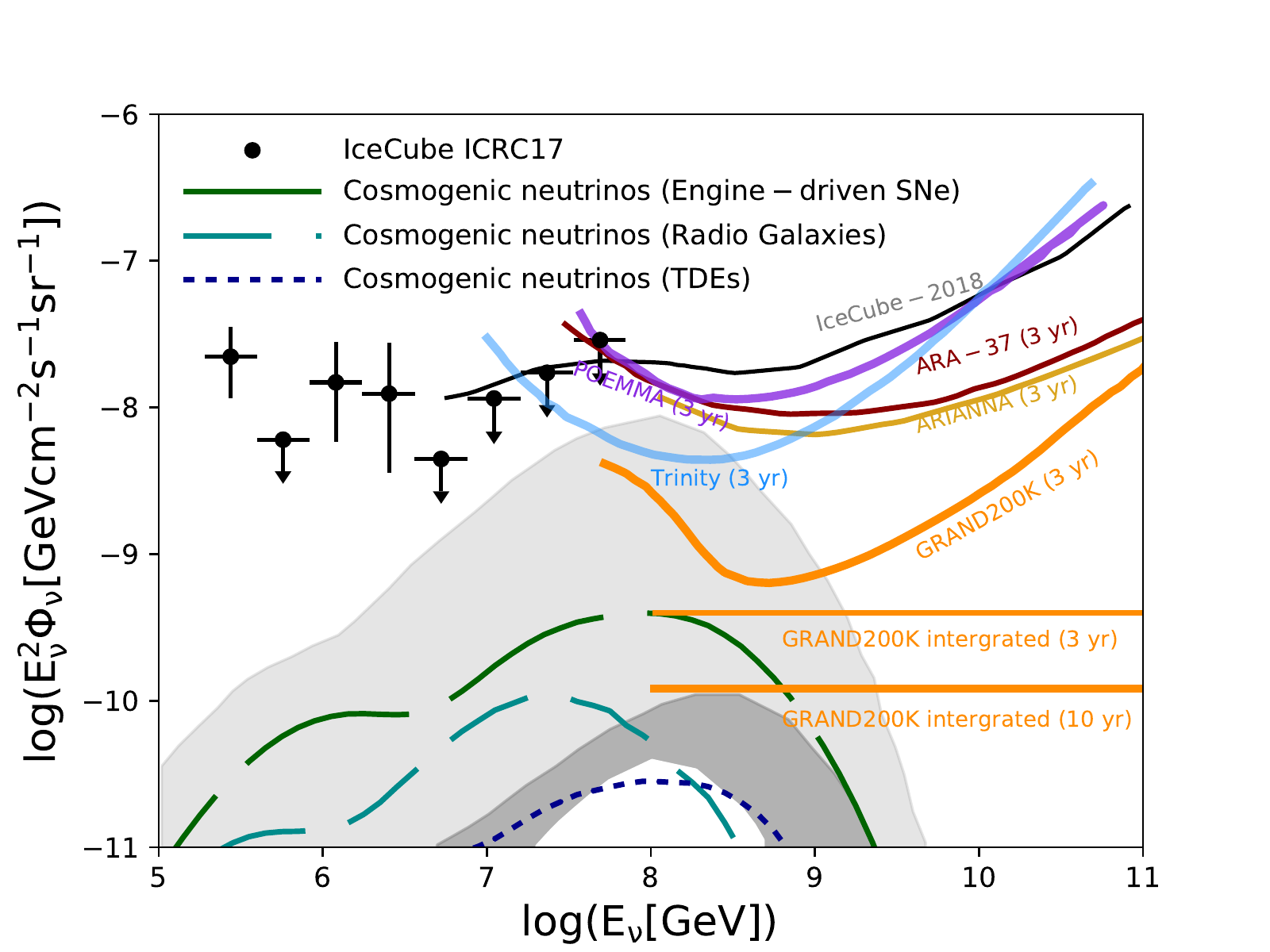}
\caption{All-flavor cosmogenic neutrino fluxes predicted from different UHECR nuclei sources and detection sensitivities for various experiments. The gray-shaded region is taken from the results of Ref.~\cite{AlvesBatista:2018zui}. The black solid line is the 90\% CL upper limit derived from IceCube~\cite{Aartsen:2018vtx}. The brown and yellow line is the 90\% CL differential upper limit derived from ARA-37~\cite{Allison:2015eky} and ARIANNA~\cite{Barwick:2014pca}, respectively. The blueviolet line is the differential upper limit from POEMMA~\cite{Olinto:2017xbi, POEMMA_UHECR2018:Talk}. The dodgerblue line is the differential upper limit from Trinity~\cite{Otte:2018uxj}. The orange line is the 90\% CL differential upper limit (solid) and integral upper limit (dashed) derived from GRAND~\cite{Alvarez-Muniz:2018bhp}.
 \label{fig:cosmogenic_neutrino}}
\end{figure}

The flux of cosmogenic neutrinos can be estimated using the following formula, 
\begin{eqnarray}
\Phi_\nu(E_\nu) &\equiv& \frac{dN_\nu}{dE_\nu dA d\Omega dt} \nonumber \\ &=& \sum_{A^\prime} \frac{c}{4\pi H_0} \int_{z_{\rm min}}^{z_{\rm max}}dz \frac{F(z) \dot{\rho}_0}{(1 + z) \sqrt{\Omega_M (1+z)^3 + \Omega_\Lambda}} \nonumber \\ &\times & \int_{E^\prime_{\rm min}}^{E^\prime_{\rm max}} dE^\prime \frac{dN_{A^\prime}}{dE^\prime}\frac{d\eta_{A^\prime \nu}(E^\prime, E_\nu, z)}{dE_\nu},\,\,\,\,\,\,\,\,\,\,
\end{eqnarray}
where $\eta_{A^\prime \nu}$ is the neutrino yield function which reflects the fraction of neutrinos with energy $E_\nu$ originated from nuclei with mass number $A^\prime$ and energy $E^\prime$ at redshift $z$.
The value of $\eta_{A^\prime \nu}$ can be estimated numerically using CRPropa 3~\cite{Batista:2016yrx}, where UHECR nuclei are propagated through the universe as in Sec.~\ref{sec:three_B}.
We show the results in Fig.~\ref{fig:cosmogenic_neutrino} where the flux of cosmogenic neutrinos (green line) can reach a level of a few $\times 10^{-10} \rm~GeV~cm^{-2}~s^{-1}~sr^{-1}$ and have peak energy around $0.1 - 1 \rm~EeV$. Note the small bump appeared in the lower energy range $\sim \rm~PeV$ is due to the effect of neutron beta decay. We can see that the planned neutrino detector GRAND can reach the required sensitivity to detect cosmogenic neutrinos predicted in this work after $\sim 10\rm~yr$ observations~\cite{Alvarez-Muniz:2018bhp}.
For comparison, we also show the energy spectrum of cosmogenic neutrinos predicted from different candidate sources, including radio galaxies in the shear acceleration scenario (cyan dashed line)~\cite{Kimura:2017ubz} and tidal disruption events (TDEs) where a white dwarf is disrupted by an intermediate mass black hole (blue dotted line)~\cite{Zhang:2017hom}. The flux of cosmogenic neutrinos is sensitive to the source redshift evolution because most of the detected neutrinos are produced at redshift $z \sim 1$, while the observed UHECR nuclei mainly originated from sources within the local universe $\leq 100\rm~Mpc$. Engine-driven SNe have a fast redshift evolution, which traces the star formation history. On the other hand, the number density of TDEs usually has a negative redshift evolution~\cite{Sun:2015bda}.

\begin{figure}
\includegraphics[width=\linewidth]{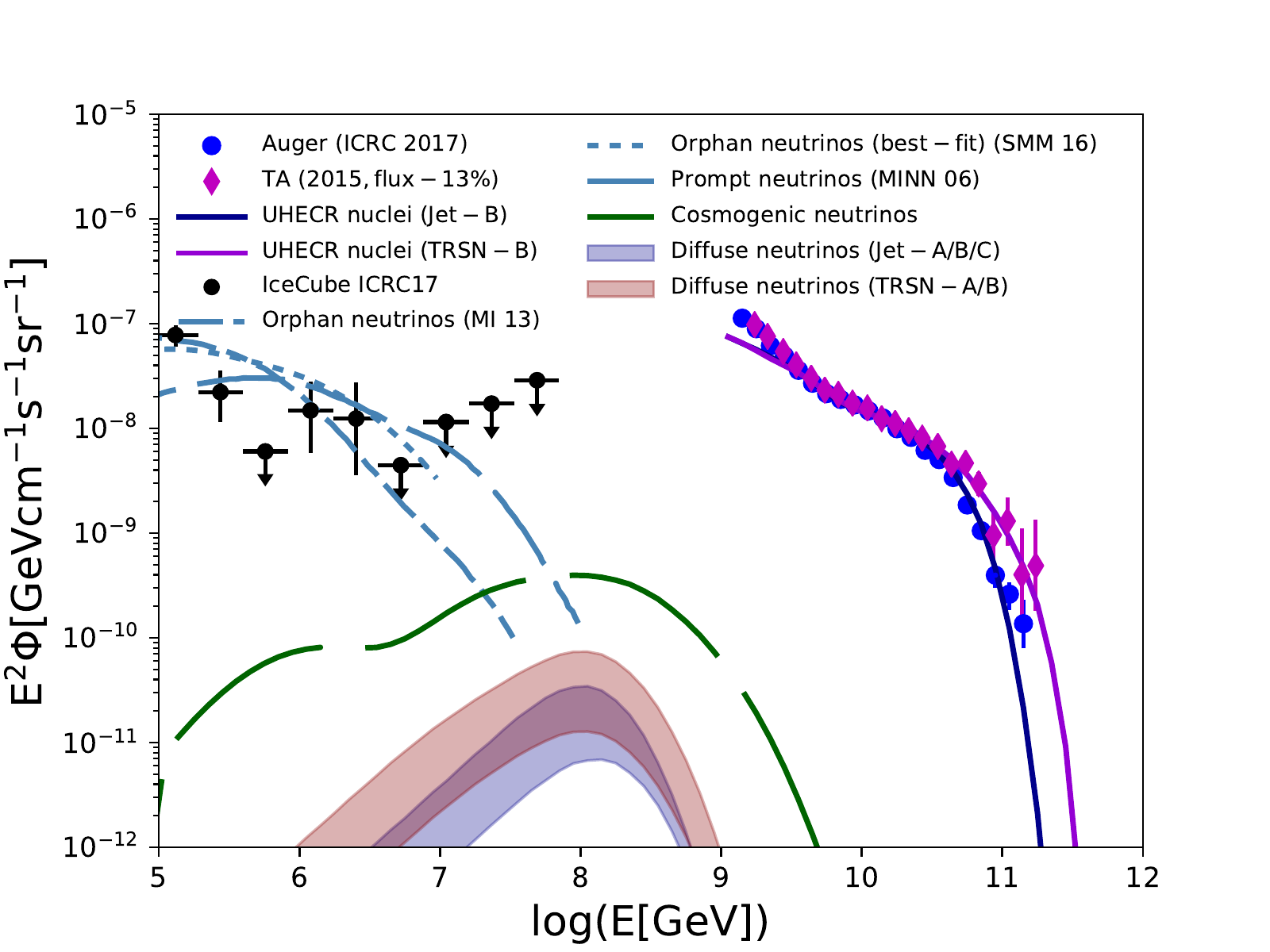}
\caption{Energy spectra of UHECR nuclei and diffuse neutrinos from engine-driven SNe or LL GRBs. We also show diffuse orphan neutrinos in the case of choked jets from Ref.~\cite{Murase:2013ffa} (dot-dashed line) and Ref.~\cite{Senno:2015tsn} (dotted line), and LL GRB prompt neutrinos (dashed line)~\cite{Murase:2006mm, Murase:2015xka}, respectively. 
The observation data of UHECR nuclei (brown triangle) are taken from Ref.~\cite{Aab:2017njo} and neutrinos are measured by IceCube~\cite{Aartsen:2017mau}. 
%The brown solid line are the best-fit UHECR nuclei energy spectrum assuming Si-R \RN{2} composition in the completely survival regime~\cite{Zhang:2017moz}.
%Bottom panel: the distribution of $\langle X_{\rm max}\rangle$ and $\sigma(X_{\rm max})$~\cite{Zhang:2017moz} and the observation data are taken from Ref.~\cite{Aab:2017njo}. 
\label{fig:xmax_combine}}
\end{figure}

The main results derived in this work are summarized in Fig.~\ref{fig:xmax_combine}, where the predicted energy spectra of UHECR nuclei and neutrinos from engine-driven SNe are shown together. 

\section{\label{sec:five} Discussion}

\subsection{Common origin of IceCube neutrinos and UHECRs?}
The flux of the observed diffuse neutrinos within energy range from $\sim 100\rm~TeV$ to a few PeV by IceCube~\cite{Aartsen:2014gkd} is comparable to the Waxman-Bahcall (WB) bound for a spectral index of $s=2.0$~\cite{Waxman:1998yy} and the nucleus-survival bound for a spectral index of $s\sim2.2-2.3$~\cite{Murase:2010gj}. The fact that the diffuse energy fluxes of all three messengers are similar has led to the development of the unification picture~\cite{Murase:2016gly,Fang:2017zjf}.

The flux of diffuse source neutrinos predicted in this work is $\sim 2 - 3$ orders lower than the observed flux of IceCube TeV-PeV neutrinos. 
However, it is quite plausible that UHECR nuclei and TeV-PeV neutrinos are produced at different regions within the same source. 
For example, in the hybrid ``two-zone'' model~\cite{Zhang:2017moz}, it has been suggested that UHECR nuclei mainly come from larger radii, where the survival is easy, whereas TeV-PeV neutrinos are produced efficiently near or under the photosphere~\cite{Murase:2006mm,Kashiyama:2012zn,Murase:2013ffa,Senno:2015tsn}. Indeed, the mechanism of prompt emission from LL GRBs may be different from that of canonical high-luminosity GRBs. The most popular explanation is shock breakout emission of transrelativistic SNe~\cite{Campana:2006qe,Nakar:2015tma}. Transrelativistic SNe naturally originate from choked jets~\cite{Nakar:2015tma,Senno:2015tsn}, and high-energy neutrino emission from low-power choked jets can explain the IceCube data even in the 10-100~TeV range~\cite{Murase:2013ffa,Senno:2015tsn}. Our RS model studied in this work is consistent with this shock breakout scenario, and the common origin requires the two-zone picture.  Note that our model is also consistent with late time observations at optical wavelengths. 
As earlier calculated by Ref.~\cite{Murase:2006mm}, the successful jet scenario~\cite{Toma:2006iu,Ghisellini:2006ng} may also be viable, in which prompt emission comes from the jet that breaks out of the star.  Following this specific scenario, a recent paper by Ref.~\cite{Boncioli:2018lrv} attempted to simultaneously explain the PeV neutrino flux and UHECR data. 

\subsection{Maximum acceleration energy?}
We showed that the maximum energy of UHECR nuclei reaches $\sim{\rm a~few}\times(Z / 26) 10^{20}\rm~eV$ in the framework of RS formed by both jets and transrelativistic ejecta. 
We confirmed the findings of the original papers~\cite{Murase:2006mm, Murase:2008mr}, which found that the maximum energy can be as high as $\sim{\rm a~few}\times10^{20}$~eV for LL GRBs like GRB 06218 without violating observational data (see also Refs.~\cite{Liu:2011cua, Boncioli:2018lrv, Zhang:2017moz}).   
A recent paper by Ref.~\cite{Samuelsson:2018fan} claimed that it may be difficult for protons or iron nuclei to reach the maximum acceleration energy $\sim 10^{20}\rm~eV$, if one considers the synchrotron modeling of observed prompt emission for GRBs. 
Note that even in that work, if one chooses the fiducial parameter sets based on the observations of GRB 060218, $L_\gamma \sim10^{47}\rm~erg~s^{-1}$ and $\Gamma \sim1-10$, the maximum energy can exceed $\sim 10^{20}\rm~eV$.  Also, the mechanism of prompt emission from LL GRBs is yet uncertain, and the constraints on synchrotron emission do not directly applied in the shock breakout scenario. 
%For HL GRBs, the problem is even serious because the survival of UHECR nuclei requires larger radii, e.g., $\sim 10^{15}\rm~cm$ where the magnetic field strength is too low to reconcile with the observed peak energy of gamma-ray emission~\cite{Zhang:2017moz}.
%\subsection{\btz{Constraints from the electromagnetic observations}}
%\btz{When we modeling the emission from external reverse-forward shock, one important factor is $f_e$, which denotes the fraction of electrons that are injected into the acceleration process, and the remained electrons can be treated as thermal electrons occupied ($1 - f_e$) of total electrons. It have been shown that the thermal electron emission can play an import role in the early stage of the evolution of reverse-forward shock~\cite{Eichler:2005ug, Giannios:2009df, Ressler:2017qjo}. Assuming the thermal electrons have peak energy $\eta \Gamma m_e c^2$, the ratio between the injection frequency of thermal electrons and non-thermal electrons is $(\eta m_e / m_p)^2$.}

\section{\label{sec:six}Summary}
In this work, we investigated UHECR nuclei and neutrinos originating from engine-driven SNe. 
In Sec.~\ref{sec:two}, we showed that the acceleration and survival of UHECR nuclei is possible by RS formed by both jets and transrelativistic ejecta. We considered five models, Jet-A/B/C and TRSN-A/B, which are different by the composition and survival regime of UHECR nuclei. 
The models Jet-A/B/C have a composition similar to Si-R \RN{1} in Ref.~\cite{Zhang:2017moz} and models TRSN-A/B have a hypernova composition with heavier iron-group nuclei. 
Both of the models Jet-B and TRSN-B are in the complete survival regime, and we found Jet-B is compatible with the Auger data, including the energy spectrum and composition. 
%\btz{where the details of the survival may be affected by the ambient density.} 
%We found Jet-B is compatible with the Auger data, including the energy spectrum and composition. 
On the other hand, model TRSN-B can give a better fit to the TA data. 
It implies that the flux discrepancy between Auger and TA data around the suppression energy could in principle be caused by the enhanced number of engine-driven SNe in the northern sky. However, we should keep in mind that the survival of nuclei is sensitive to the ambient density, as in model Jet-A/C and TRSN-A, which may affect the composition of escaped UHECR nuclei.
In Sec.~\ref{sec:three}, we estimated neutrinos that are coproduced with UHECR nuclei. At first, we calculated the received neutrino fluences at Earth from a single engine-driven SN, which is located at distance $z = 0.005$. Then we found that the detection is challenging even for planned neutrino detectors, such as GRAND. We also showed that the neutrino fluences can be $\sim 1$ orders higher in the partial survival regime than in the complete survive regime. In the partial survival regime, neutrinos produced from secondary protons/neutrons can play a dominant role in the high-energy range. 
Because of the relatively high event rate of engine-driven SNe compared to classical GRBs, the diffuse neutrinos can reach a flux level of $\sim 10^{-11} - 10^{-10} \rm~GeV~cm^{-2}~s^{-1}~sr^{-1}$. This cumulative flux can be detected by GRAND after $\sim 10$ years observations. We found that the flux of cosmogenic neutrinos is also comparable or even higher than the diffuse source neutrinos. This means that UHECR nuclei may lose most of their energy in the intergalactic space other than inside the source.
Finally, we discuss the common origin of IceCube TeV-PeV neutrinos and UHECR nuclei. Our results confirmed Ref.~\cite{Zhang:2017moz}, which suggested that the hybrid ``two-zone'' model may give an alternative explanation of UHECR nuclei and IceCube TeV-PeV neutrinos. This picture is consistent with the most popular interpretation that the prompt emission of LL GRBs originates from shock breakout of transrelativistic SNe in a dense stellar wind.

\medskip
\begin{acknowledgments}
The work of K. M. is supported by Alfred P. Sloan Foundation and the U.S. National Science Foundation (NSF) under grants NSF Grant No. PHY-1620777. B.T.Z. is supported by High-performance Computing Platform of Peking University.
The common origin of IceCube neutrinos and UHECRs are discussed in Ref.~\cite{Zhang:2017moz}. Then, while we are working on this manuscript, related papers came out~\cite{Boncioli:2018lrv, Samuelsson:2018fan}.  One of the differences is that this work focuses on the external RS model, in which prompt emission can be attributed to shock breakout emission of transrelativistic SNe. Our model prediction is also consistent with both of the optical and x-ray data observed for GRB 060218.
\end{acknowledgments}

\appendix

\bibliography{bzhang}% Produces the bibliography via BibTeX.

\end{document}